\documentclass[12pt,preprint]{aastex}
\usepackage{natbib}
\newcommand \zaz          {{$z_a\kern -1.5pt \approx\kern -1.5pt z_e$}}
\newcommand \zllz         {{$z_a\kern -3pt \ll\kern -3pt z_e$}}

\newcommand{\halpha}{\ifmmode {\rm H}\alpha \else H$\alpha$\fi}
\newcommand{\hbeta}{\ifmmode {\rm H}\beta \else H$\beta$\fi}
\newcommand{\hgamma}{\ifmmode {\rm H}\gamma \else H$\gamma$\fi}
\newcommand{\hdelta}{\ifmmode {\rm H}\delta \else H$\delta$\fi}
\newcommand{\lya}{\ifmmode {\rm Ly}\alpha \else Ly$\alpha$\fi}
\newcommand{\lyb}{\ifmmode {\rm Ly}\beta \else Ly$\beta$\fi}
\newcommand{\heI}{\ifmmode {\rm He}\,{\sc i}\,\lambda5876 \else 
	          He\,{\sc i}\,$\lambda5876$\fi}
\newcommand{\heII}{\ifmmode {\rm He}\,{\sc ii}\,\lambda4686 \else 
	           He\,{\sc ii}\,$\lambda4686$\fi}

\newcommand{\heii}{He\,{\sc ii}}
\newcommand{\fe}{Fe}
\newcommand{\feii}{Fe\,{\sc ii}}

\newcommand{\cii}{C\,{\sc ii}}
\newcommand{\ciii}{\ifmmode {\rm C}\,{\sc iii} \else C\,{\sc iii}\fi}
\newcommand{\civ}{\ifmmode {\rm C}\,{\sc iv} \else C\,{\sc iv}\fi}

\newcommand{\niii}{N\,{\sc iii}}
\newcommand{\niv}{N\,{\sc iv}}
\newcommand{\nv}{N\,{\sc v}}
\newcommand{\oi}{O\,{\sc i}}

\newcommand{\oiii}{O\,{\sc iii}}
\newcommand{\ob}{[O\,{\sc iii}]\,$\lambda5007$}

\newcommand{\ov}{O\,{\sc v}}
\newcommand{\ovi}{O\,{\sc vi}}

\newcommand{\mgii}{Mg\,{\sc ii}}
\newcommand{\siiv}{Si\,{\sc iv}}

\newcommand{\sizwo}{Si\,{\sc ii}}
\newcommand{\siiii}{Si\,{\sc iii}}

\newcommand{\aliii}{Al\,{\sc iii}}

\slugcomment{To be published in Astrophysical Journal}

\shorttitle{Continuum and Emission Line Strength Relations}
\shortauthors{Dietrich et al.}

\begin{document}

\title{Continuum and Emission Line Strength Relations for a large
       Active Galactic Nuclei Sample}


\author{M.\,Dietrich\altaffilmark{1} and
        F.\,Hamann\altaffilmark{1}}
\affil{Department of Astronomy, University of Florida, 211 Bryant Space 
       Science Center, Gainesville, FL 32611-2055, USA}
%\and
% 
\author{J.C.\,Shields\altaffilmark{2} and
        A.\,Constantin\altaffilmark{2}}
\affil{Department of Physics \& Astronomy, Ohio University, Clippinger 
       Research Labs.251B, Athens, OH 45701-2979, USA}
%\and
%
\author{M.\,Vestergaard\altaffilmark{3}}
\affil{Department of Astronomy, The Ohio State University, 140 West 18th Av.,
       Columbus, OH 43210-1173, USA.}
%\and
%
\author{F. Chaffee\altaffilmark{4}}
\affil{W.M. Keck Observatory, 65-1120 Mamalahoa Hwy., Kamuela, HI 96743, USA}

%\and
%
\author{C.B.\,Foltz\altaffilmark{5}}
\affil{Multiple Mirror Telescope Observatory, University of Arizona, Tucson,
       AZ 85721, USA}
\and
\author{V.T.\,Junkkarinen\altaffilmark{6}}
\affil{Center for Astrophysics and Space Sciences, Code 0424, 9500 Gilman 
    Drive, University of California at San Diego, La Jolla, CA 92093-0424, USA}

\email{dietrich@astro.ufl.edu}
%\email{aastex-help@aas.org}
%
%\and
%
%\author{R. J. Hanisch\altaffilmark{5}}
%\affil{Space Telescope Science Institute, Baltimore, MD 21218}

%\altaffiltext{1}{Visiting Astronomer, Cerro Tololo Inter-American Observatory.
%CTIO is operated by AURA, Inc.\ under contract to the National Science
%\altaffiltext{2}{Society of Fellows, Harvard University.}
%\altaffiltext{3}{present address: Center for Astrophysics,
%    60 Garden Street, Cambridge, MA 02138}
%\altaffiltext{4}{Visiting Programmer, Space Telescope Science Institute}
%\altaffiltext{5}{Patron, Alonso's Bar and Grill}


\begin{abstract}
We report on the analysis of a large sample of 744 type 1 Active Galactic
Nuclei, including quasars and Seyfert\,1 galaxies across the redshift
range from $0\la z \la 5$ and spanning nearly 6 orders of magnitude in 
continuum luminosity.
We discuss correlations of continuum and emission line properties 
in the rest-frame ultraviolet and optical spectral ranges.
The well established Baldwin Effect is detected for almost all emission 
lines from \ovi $\lambda 1034$ to \ob .
Their equivalent widths are significantly anti-correlated with the continuum 
strength, while they are nearly independent of redshift.
This is the well known Baldwin Effect. Its slope $\beta $, measured as 
log\,$\,W_\lambda \propto \beta ~{\rm log}\,\lambda 
L_\lambda (1450 {\rm \AA })$,
shows a tendency to become steeper towards higher luminosity.
The slope of the Baldwin Effect also increases with the ionization energy 
needed to create the individual lines.
In contrast to this general trend, the \nv $\lambda 1240$ equivalent width is 
nearly independent of continuum luminosity and remains nearly constant. 
The overall line behaviors are consistent with softer UV continuum shapes 
and perhaps increasing gas metallicity in more luminous Active Galactic
Nuclei.
\end{abstract}




\section{Introduction}

Broad emission lines (BELs) are a defining property of quasar spectra. 
Nearly 25 years ago Baldwin (1977) discovered an anti-correlation between  
the equivalent width in \civ $\lambda 1549$, $W_\lambda$(\civ ), 
and the continuum luminosity, $L_\lambda$(1450 \AA ), measured at 1450 \AA\
in the quasar rest-frame.
This result was based on a sample of 20 quasars spanning $\sim2$ orders of 
magnitude in continuum luminosity in the redshift range $1.24 \la z\la 3.53$. 
It has become known as the ``Baldwin Effect'' (hereafter BEff). 
The BEff was subsequently confirmed in \civ $\lambda 1549$, 
as well as other BELs such as \lya\ and \ovi $\lambda 1034$ 
(V\'eron-Cetty, V\'eron, \& Tarenghi 1983; Baldwin, Wampler, \& Gaskell 1989; 
 Kinney, Rivolo, \& Koratkar 1990; Osmer, Porter, \& Green 1994; 
 Zheng, Kriss, \& Davidsen 1995; Green, Forster, \& Kuraszkiewicz 2001; 
 see Osmer \& Shields 1999 for a recent review).

A major impetus for studying the BEff was that it might be useful for
calibrating Active Galactic Nuclei (AGNs) luminosities, e.g. based on 
$W_\lambda$(\civ ).
The AGNs could then be used as cosmological standard candles. 
But in the following years studies of bigger quasar samples revealed a large 
scatter in the anti-correlation of the continuum luminosity vs. emission line 
strength (Baldwin, Wampler, \& Gaskell 1989; Zamorani et al.\,1992).
The BEff is nonetheless important as a diagnostic of AGN structure and, 
perhaps, metal abundances (Korista, Baldwin, \& Ferland 1998).
The relation between the continuum luminosity and the relative emission line 
strengths and ratios can be used to study the evolution and physics of the 
quasar phenomenon (Baldwin 1999).
In particular, this correlation can be used to test model predictions for the 
dependence of the shape of the continuum spectral energy distribution as a 
function of luminosity (Binette et al.\,1989; Netzer, Laor, \& Gondhalekar 
1992; Zheng \& Malkan 1993; Wandel 1999a,b).

The most fundamental problem, however, is that the physical cause of the BEff
remains unknown. 
It was suggested by Mushotzky \& Ferland (1984) that the observed relation
can be explained by an anti-correlation of the ionization parameter, U, and 
the continuum luminosity, with $U = Q(H) / 4 \pi r^2 c n_H$, where Q(H) is the
number of hydrogen ionizing photons emitted per second by the central continuum
source, r is the distance between the continuum source and the emission line
region, and $n_H$ is the hydrogen density in the line-emitting cloud.
Assuming an additional relation of decreasing covering factor with increasing 
continuum strength, the observed BEff could be well described for 
\civ $\lambda 1549$.
However, this model does not naturally explain the BEff in the full range of 
measured lines. It predicts, for example, a lack of a BEff for Ly$\alpha$ even
though it is clearly detected (e.g., Kinney, Rivolo, \& Koratkar 1990; 
Osmer, Porter, \& Green 1994; Laor et al.\,1995; Green 1996).

Another important clue to the physical cause of the BEff is that the strength 
of the relationship (i.e., the slope of the $W_\lambda $ -- $L_c$ 
anti-correlation) seems to depend on the ionization energy of the 
emission-line species (Zheng et al.\,1995; Espey \& Andreadis 1999). 
In particular, the equivalent widths of high-ionization lines like 
\ovi $\lambda 1034$ decrease more dramatically with $L_c$ than lines with 
moderate ionization energies like \civ $\lambda 1549$ or low ionization 
energies like \mgii $\lambda 2798$ and Balmer emission lines. 
Our results in the present paper confirm this claim, and improve upon the 
overall empirical characterization of the BEff correlations. 

The BEff might result from a more fundamental correlation between the continuum
luminosity, $L_c$, and the shape of the ionizing (EUV\,--\,soft X-ray)
continuum (Binette et al.\,1989; Zheng \& Malkan 1993; Zheng et al.\,1997; 
Korista et al.\,1998). Netzer (1985, 1987) and Netzer et al.\,(1992) suggested 
accretion disc models to explain the observed continuum and emission line 
correlations. 
Recently, Wandel (1999\,a,b) added to this the growth in black hole mass by 
accretion in the accretion disc model. His analysis predicts that the 
continuum luminosity increases towards higher black hole mass and that the 
shape of the ionizing continuum becomes softer.
Hence, it is suggested that the BEff is driven by a softening of the ionizing
continuum towards higher luminosities. 
This model is attractive because the UV\,--\,X-ray spectral softening has been
well documented by observations (Tananbaum et al.\,1986; Wilkes et al.\,1994;
Green et al.\,1995), and because the spectral softening provides a natural
explanation for steeper BEff slopes in higher ionization lines.

An additional possibility is that the BEff is driven, at least in part, by a 
trend for higher metallicities in more luminous AGNs. Korista et al.\,(1998) 
presented theoretical results showing the dependence of the emission line 
equivalent widths on both the continuum shape and the metallicity of the gas.
Their proposal is based on evidence for higher metal abundances in more 
luminous quasars (Hamann \& Ferland 1993, 1999; Dietrich et al. 1999; 
Dietrich \& Wilhelm-Erkens 2000), and on the suggestion that more luminous 
quasars reside in more massive host galaxies, which will naturally be more 
metal rich (Cen \& Ostriker 1999; Pettini 1999; Kauffmann \& Haehnelt 2000; 
Granato et al.\,2001; see Hamann \& Ferland 1999 for a review). 

Our combination of ground-based and satellite data also provides the 
opportunity to study evolutionary aspects of the line strengths in detail --- 
e.g. by measuring the same rest-frame UV lines across a wide range of 
redshifts. 
One serious complication affecting previous work is that luminosity and 
redshift are often correlated in quasars samples, because the quasars are 
selected from magnitude-limited surveys. Most studies have supported the 
original claim by Baldwin (1977), that W$_\lambda$ scales inversely with 
luminosity and that there is no significant trend with redshift $z$ 
(e.g., Kinney, Rivolo, \& Koratkar 1990;  Osmer et al.\,1994; 
Francis \& Koratkar 1995; Wilkes et al.\,1999). 
However, some recent work based on the Large Bright Quasar Survey (LBQS) 
claims that the relationship to redshift is even stronger than with luminosity 
(Forster et al.\,2001; Green, Forster, \& Kuraszkiewicz 2001). 

The present paper is the first in a series in which we examine the 
emission-line properties in a large sample of $744$ type 1 AGNs.
The sample includes Seyfert\,1 galaxies and both radio-loud and radio-quiet 
quasars spanning an unprecedented wide range in both redshift ($0\la z\la 5$) 
and intrinsic luminosity ($\sim 6$ orders of magnitude). 
The database and data processing will be described in detail in a future paper.
A major concern regarding BEff studies is that selection effects in AGN 
samples, e.g. pertaining to emission-line strengths, might bias some 
measurements of the BEff. Our sample has the advantage of being drawn from 
a variety of surveys, with objects selected based on radio properties, 
grism spectroscopy, or broad-band color criteria. Hence, no specific selection
criteria were applied for the data. 
Another key advantage is that our sample includes a wide range of luminosities
at various redshifts. We can therefore address the separate redshift and
luminosity dependences in the emission line data.

There are several interesting problems we wish to address. The most basic issue
is to quantify the nature of empirical correlations with $L_c$ and/or $z$.
In an upcoming study we will further examine correlations with other 
AGN properties such as radio-loudness (Baldwin 1977; Sargent, Steidel, \&
Boksenberg 1989;
Steidel \& Sargent 1991; Francis \& Koratkar 1995). We will also report on
trends with $L_c$ or $z$ among various emission-line metallicity indicators.
Here we present an initial analysis on the nature of the BEff. 
Our approach is to construct composite spectra for specific $L_c$ and $z$
intervals, thus providing high signal-to-noise spectra and allowing us to 
study trends in both weak and strong emission lines.

The AGN sample is described briefly in \S2\ and the calculation and analysis
of the composite spectra in \S3 and \S4.
The main results appear in several figures in \S5. 
We compare our results with prior studies and discuss them in the context
of suggested models in \S6. 
Throughout this paper we use the cosmological parameters  
H$_o = 65$ km\,s$^{-1}$\,Mpc$^{-1}$, $\Omega _M = 0.3$, and
$\Omega _\Lambda = 0$ (Carroll, Press, \& Turner 1992).
Introducing $\Omega_\lambda = 0.7$ (Netterfield et al.\,2002) instead of
$\Omega_\lambda = 0.0$, the luminosities at the highest redshifts would
be $\sim 10$\,\%\ smaller, while at low redshifts they would reach a
maximum of $\sim 20$\,\%\ larger at redshift $z\simeq 1$.

\section{The Quasar Sample}

We have compiled a large sample of rest-frame visible and ultraviolet spectra
for $826$ type 1 AGNs. A majority of the spectra were obtained by several 
groups for different studies over the last 20 years using ground-based 
instruments as well as 
{\it International Ultraviolet Explorer (IUE)} 
and {\it Hubble Space Telescope (HST)}
(Bahcall et al.\,1993; 
 Baldwin et al.\,1989;
 Chaffee et al.\,1991;
 Corbin \& Boroson 1996;
 Kinney et al.\,1991; 
 Lanzetta, Turnshek, \& Sandoval 1993;
 Laor et al.\,1995;
 Sargent et al.\,1988,\,1989;
 Schneider, Schmidt, \& Gunn 1991a,\,b;
 Steidel 1990;
 Steidel \& Sargent 1991;
 Steidel unpublished;
 Storrie-Lombardi et al.\,1996;
 Weymann et al.\,1991,\,1998;
 Wills et al.\,1995; 
 Zheng et al.\,1997). 
In total, the sample contains 351 quasar spectra measured with {\it HST}.
Furthermore, we observed a large number of the quasars in our sample at 
redshifts $z\ga 3$ 
(Constantin et al.\,2002;
 Dietrich et al.\,1999;
 Dietrich \& Wilhelm-Erkens 2000; 
 Dietrich et al.\,2002a,\,b).
For this study we exclude Broad-Absorption Line quasars (BAL\,QSOs), although 
there are indications that their emission line properties do not differ from 
non-BAL quasars (Weymann et al.\,1991).
We also excluded several quasar spectra with extremely poor signal-to-noise 
ratios. The sample we investigate consists of $744$ type 1 AGNs.

Because this AGN sample was compiled from many independent projects, it
encompasses a wide range of selection criteria.
Most of the quasars especially at high redshifts were discovered by color 
selection techniques (e.g., Storrie-Lombardi et al.\,1996) 
or found by objective prism surveys 
(e.g., Chaffee et al.\,1991; Sargent et al.\,1988,\,1989; 
       Schneider at al.\,1991a,\,b).

We used the radio flux densities given in V\'eron-Cetty \& V\'eron (2001) to
determine the radio loudness for the quasars. 
Following Kellermann et al.\,(1989) we calculated the parameter 
$R_L = \log (F_R / F_B)$ with $F_R$ as the radio flux density at $5$\,GHz
and $F_B$ as the B-band flux, each in the rest frame. Our classifications of 
radio-loud quasars are consistent with classifications available in the 
literature (Wills et al.\,1995; Bischof \& Becker 1997; Wilkes et al.\,1999; 
Stern et al.\,2000). In total our sample contains $265$ radio-loud quasars 
and $479$ radio-quiet AGNs.  

Kinney et al.\,(1990) and Pogge \& Peterson (1992) pointed out that a 
significant fraction of the scatter in the BEff in Seyfert\,1 galaxies is 
caused by variability. 
They found that observations obtained at multiple epochs for individual
sources display a BEff-like correlation (the {\it intrinsic} BEff), but with a
significantly steeper slope than the BEff observed for an ensemble of AGNs. 
The physical mechanisms which drive each effect might be different, i.e., 
the ensemble BEff might be due primarily to variations in black hole mass 
while the intrinsic BEff might be caused by changes of the accretion rate 
$\dot M$ (Green 1999).
For $8$ of the $36$ low luminosity AGNs 
(log\,$\lambda L_\lambda (1450{\rm \AA }) \leq 44$, for $\lambda L_\lambda$ in
erg\,s$^{-1}$), the spectra are means from intense monitoring campaigns 
(Crenshaw et al.\,1996; Collier et al.\,1998,2001; Dietrich et al.\,1993;
 Goad et al.\,1999; Kaspi et al.\,1996; Korista et al.\,1995;
 Peterson et al.\,1994; Reichert et al.\,1994; Santos-Lle\'o et al.\,2001;
 Stirpe et al.\,1994; Wanders et al.\,1997).
At the high luminosity end of the luminosity distribution, it has been shown 
that quasars vary on longer timescales and with smaller amplitudes
(e.g., Kaspi et al.\,2000). 
For these reasons, the scatter caused by intrinsic variability is not
expected to strongly influence our analysis of the present large quasar 
sample, particularly when comparisons of composite spectra are employed.

We transformed each quasar spectrum to the rest-frame using a redshift 
measured from \civ $\lambda 1549$. To determine the redshift we fit a Gaussian
profile to the upper part of the \civ $\lambda 1549$ emission line having 
$I_\lambda \geq 50$\,\%\ of the peak intensity. 
We corrected each \mbox{AGN} spectrum for Galactic extinction (Seaton 1979), 
based on the neutral hydrogen column density $N_H$ expressed in units of 
$10^{20}$\,cm$^{-2}$, assuming $E_{B-V} = N_H / 60$ (Dickey \& Lockman 1990).

The redshift distribution of the quasar sample is presented in Figure 1 as a 
function of continuum luminosity, $\lambda L_\lambda (1450{\rm \AA })$.
The quasars cover a redshift range of $0\la z \la 5$ and nearly $6$ orders of 
magnitude in luminosity. 
For most of the redshift range a reasonable coverage in luminosity is achieved,
spanning at least $3$ orders of magnitude for $z\ga 1$.
The radio-loud and radio-quiet quasars are displayed with different symbols 
in Figure 1. Their distributions across luminosity and redshift are very 
similar within out sample, although the fraction of radio-loud quasars 
increases for redshifts $z\la 2$.

\section{Composite Spectra}

A common approach to study the BEff is to measure the emission line 
properties of individual quasars.
However, the results are often limited especially for the weaker emission 
lines by the low signal-to-noise ratio of individual spectra.
There can also be significant problems introduced by including upper
limits on weak lines in individual spectra.
To avoid those problems, we computed composite spectra in different intervals
of redshift and luminosity.
The analysis of these spectra has the advantage of more clearly representing
the average properties of the sample used to calculate each composite. 
Figure 1 indicates the ranges in luminosity and redshift used to compute the 
composites. Except for the low and high luminosity ends of our sample, each 
composite spectrum is based on more than $20$ individual quasar spectra.
The scatter introduced by different individual spectra is taken into account
by calculating a root-mean-square (standard deviation) spectrum for each 
composite, and using that to estimate the uncertainties in the measured 
equivalent widths (\S4).

Before combining the individual quasar spectra, we normalized the flux 
densities to unity in a $20$\,\AA\ wide continuum range centered at 
$\lambda = 1450$\,\AA\ by multiplying each spectrum with a suitable factor.
We used these normalized quasar spectra to calculate the mean and rms spectrum
for each composite, and the number of spectra contributing in each wavelength
bin. 

To avoid the influence of strong narrow absorption lines, we developed a 
search routine to detect strong narrow absorption features and 
exclude them from the calculation of the composite spectrum.
To search for strong absorption lines we applied the following method (a more 
detailed description will be given in a subsequent paper on the properties of 
the entire quasar sample, Dietrich et al.\,2002c).
First, a preliminary mean spectrum is calculated for the sample of spectra 
under study. 
We then divided each individual spectrum by this preliminary mean to derive a 
ratio spectrum.
Each ratio spectrum is smoothed with a running boxcar function, which provides 
in addition to the mean also an rms value for each wavelength box.
The resulting rms spectrum indicates the scatter in flux for each wavelength 
element in the immediate spectral region. A narrow absorption line will cause 
a sharp increase in the rms spectrum.
The contaminated spectral regions of individual spectra, identified by these 
rms spikes, are excluded from the calculation of the final composite spectrum,
and no interpolation has been applied.
Multiple tests with different AGNs and different sample sizes show that this 
technique sucessfully removes strong narrow absorption lines from our 
composite spectra (see below) without affecting the emission lines.
Significant effects of narrow absorption lines remain only in the Ly$\alpha $ 
forest ($\lambda \la 1216$\AA ) for composites that include quasars with 
redshift $z\ga 1$.

The resulting composite quasar spectrum for the entire sample is shown in 
Figure 2. A single power law fit to the continuum at $\lambda \geq 1200$\,\AA\ 
is also displayed; it has a slope of $\alpha = -0.43$ 
($F_\nu \propto \nu^\alpha$). This spectral index is consistent with the slope 
reported for other composite spectra (Brotherton et al.\,2001; 
Vanden\,Berk et al.\,2001; Telfer et al.\,2002).
The bottom panel of Figure 2 shows the number of quasars contributing to each 
individual wavelength element. 
For every wavelength $800 \la \lambda \la 5600$\,\AA\ this composite quasar 
spectrum is based on at least $\sim 100$ quasars.

\section{Emission Line Measurements}

To measure the emission line fluxes we analyzed each composite spectrum with 
spectral fitting software provided by MIDAS\footnote{Munich Image Data 
Analysis System, trade-mark of the European Southern Observatory}.
We employed a powerlaw continuum, an ultraviolet and optical Fe-emission 
template, a Balmer continuum 
emission template, and multiple Gaussian profiles for the broad emission lines
in multi-component fits using $\chi ^2$ minimization. 
For the ultraviolet wavelength range we used the Fe emission template which 
Vestergaard \& Wilkes (2001) carefully derived from HST spectra of I\,Zw1
(kindly provided by M.\,Vestergaard).
The primary aim was to measure the broad and prominent \feii -emission near
\mgii $\lambda 2798$. However, we also used the UV template to correct for
FeII and FeIII emission blended with \ciii ]$\lambda 1909$ and other lines 
across the wavelength range $1400 \la \lambda \la 1900$\,\AA .
For the optical wavelength region around \hbeta\ and \hgamma , we used the 
empirical \feii\ emission template obtained from I\,Zw1 (kindly provided 
by T.\,Boroson; Boroson \& Green 1992).
We adjusted the line widths in these Fe emission templates to match the 
measured widths of \civ $\lambda 1549$, \mgii $\lambda 2798$, or 
H$\beta $.

We calculated several bound-free Balmer continuum emission spectra for 
$T_e = 15000$\,K, $n_e = 10^8 - 10^{10}$\,cm$^{-3}$, and 
$0.1 \leq \tau _\nu \leq 2$ following the formalism suggested by Grandi (1982).
These Balmer continuum emission templates were supplemented for 
$\lambda > 3646$\AA\ with high order Balmer emission lines with 
$10 \leq n \leq 50$, i.e.\, H$\vartheta $ and higher (Storey \& Hummer 1995).
The individual high-order Balmer lines were represented by Gaussian profiles
and coadded.

The multi-component fit of a Balmer continuum, two \fe -emission templates,
and a power law continuum yields independent measurements of the ultraviolet 
and optical \fe -emission strengths. Furthermore, the spectral index, 
$\alpha $, of the power law continuum does not include contributions from 
Balmer continuum or from broad Fe emission.
The uncertainties associated with the fit of Balmer continuum emission and 
Fe emission templates were estimated from the $\chi ^2$ distribution for 
each fit (cf., Dietrich et al.\,2002a).

The emission line fluxes were measured after subtracting the power law 
continuum, Balmer continuum, and Fe emission spectra. The line fluxes were
determined by multi-component Gaussian fits.
In general, we fit the emission lines with a ``broad'' and a ``narrow''
Gaussian component. 
For \civ $\lambda 1549$, however, it was necessary to introduce in addition 
a very broad component 
(which turned out to have typically FWHM $\simeq 10^4$ km\,s$^{-1}$).
To determine the fluxes of Ly$\alpha$ and \nv $\lambda 1240$, which are 
severely blended, we used the components obtained from the \civ $\lambda 1549$
profile as templates (preserving their widths and redshifts). Please see
Figure 2 in Dietrich \& Hamann (2002d) for examples of fitted profiles. 
Note that most of the Ly$\alpha$ forest absorption on the blue side of the 
Ly$\alpha$ profile is corrected by the methode we applied to calculated the 
composite spectra.
The \ovi $\lambda 1034$ emission line is subject to severe corruption by 
\lya\ forest absorption.
However, in composite spectra it can be assumed that the same fraction of the 
\ovi $\lambda 1034$ emission line flux and the adjacent continuum flux is
absorbed by the \lya\ forest.
Hence, the equivalent width is not expected to be significantly altered by 
\lya\ forest absorption.
The \ovi $\lambda 1034$ emission line is also contaminated by \lyb\ emission 
in the blue wing, whose relative strength shows a wide range, 
\lyb /\ovi $\lambda 1034 = 0.30 \pm 0.21$ (Laor et al.\,1995). 
Our fits to the \ovi $\lambda 1034$ emission line profiles were achieved with 
a broad and narrow Gaussian profile, and no additional component due to 
\lyb\ emission was necessary. 

The \ciii ]$\lambda 1909$ emission line complex was reconstructed with four
gaussian profiles, corresponding to \aliii $\lambda 1857$, 
\siiii ]$\lambda 1892$, and broad and narrow Gaussian profiles for 
\ciii ]$\lambda 1909$. 
The correction for Fe emission allows a better measurement of 
\mgii $\lambda 2798$, \heii $\lambda 1640$, \niii ]$\lambda 1750$, and the 
\ciii ] emission line complex (\aliii $\lambda 1857$, \siiii ]$\lambda 1892$, 
\ciii ]$\lambda 1909$). 
The measurement of the optical \feii\ emission with an appropriate scaled
template made it possible to measure the 
\heii $\lambda 4686 $ emission line which is otherwise severely blended with 
\feii (37,38) multiplets. 

The cumulative errors in the emission line flux measurements and $W_\lambda $
were calculated as the square root of the sum of
i) the uncertainties in the $\chi ^2$ fitting and 
ii) the rms scatter that is introduced by both the noise and distribution of 
the individual spectra contributing to that particular composite spectrum,
squared.

\section{Results and Analysis}
\subsection{Trends with Redshift}

To search for a dependence of emission line properties, particularly 
$W_\lambda$, on cosmic time (redshift), we selected a narrow luminosity range.
The location of this luminosity range was also selected to achieve a comparable
number of individual quasars contributing to each redshift bin in the
$z$ -- $\lambda L_\lambda (1450{\rm \AA })$ plane. To cover a redshift range as
large as possible for an almost constant luminosity, we selected a luminosity 
range of $46.16 \leq \log\,\lambda L_\lambda (1450{\rm \AA }) \leq 47.16$
(i.e., $43.0 \leq \log\,L_\lambda (1450{\rm \AA }) \leq 44.0$).
This range contains $323$ quasars with $z\geq 0.5$ and an average luminosity of
log\,$\lambda L_\lambda (1450{\rm \AA }) = 46.67\pm 0.12$. 
We computed composite spectra for redshift bins with $\Delta z = 0.5$
centered on redshift $z = 0.75,1.25,...,3.75,4.25, {\rm and}\, 4.75$, 
respectively (Fig.\,1).
Figures 3a and 3b show the composite spectra for each redshift bin. These
composite spectra are normalized by the power law continuum which we derived 
from the multi-component fits in \S4.

As a first impression the relative line strengths in the normalized composite 
quasar spectra stay nearly constant with redshift $z$.
However, some emission lines show a marginal trend for increasing equivalent
width at the highest redshifts.
To quantify this result, we measured the equivalent width of most emission 
lines in the ultraviolet wavelength range ($\lambda \lambda 1000 - 3000$\,\AA )
and of the broad ultraviolet \feii -emission feature 
($\lambda \lambda 2200 - 3090$\,\AA ).
The results for $W_\lambda $ as a function of redshift are displayed in 
Figures 4a and 4b.
In general, the equivalent widths of the lines remain constant within 
$\sim30$\,\%.
The stronger emission lines, however, like \lya , \civ $\lambda 1549$, and 
\ciii ]$\lambda 1909$, show marginal trends for a relation of $W_\lambda $ and 
redshift. 
$W_\lambda $ stays nearly constant within less than $\sim20$\,\%\ for 
redshifts less than $z \simeq 2 - 3$, but then for $z \ga 3$ there are 
marginal indications for a slight increase of $W_\lambda $ at constant 
luminosity.
Close inspection of Figures 4a and 4b indicates that in particular the 
equivalent widths measured for the composites with $z \ga 4$ are larger 
compared to lower redshifts. 
Most of the quasar spectra in our sample with $z \ga 4$ were obtained by 
Constantin et al.\,(2002). They found evidence that in spite of strong 
contamination by \lya -forest absorption, \lya\ is stronger for quasars
with $z > 4$ than in their low-$z$ counterparts. 

\subsection{Trends with Luminosity}

Next we divided the quasar sample into eleven subsets with respect to the 
luminosity $L_\lambda (1450{\rm \AA})$. 
Except for the lowest luminosity bin, each composite spectrum covers a range 
of $\Delta \log\,\lambda L_\lambda = 0.5$ in luminosity starting at 
log\,$\lambda\,L_\lambda (1450{\rm \AA }) = 43.16$ erg\,s$^{-1}$ (Fig.\,1).
The two lowest luminosity composite spectra consist of less than 10 objects.

Figures 5a and 5b show the composite spectra ordered by luminosity for the 
ultraviolet wavelength range ($\lambda \la 2000 {\rm \AA }$), while Figure 6 
shows the longer wavelengths ($\lambda \lambda 2000 - 5600 {\rm \AA }$).
Each composite spectrum is normalized by the power law continuum fit we 
derived from the multi-component analysis (\S3).
These normalized spectra show very obviously a strong relation between the 
equivalent width, $W_\lambda $, and the continuum luminosity represented by 
$\lambda L_\lambda (1450{\rm \AA })$, i.e. the Baldwin Effect, for most of the
emission lines.
Figures 7a and 7b show the measured values of $W_\lambda $ for each 
luminosity interval.
The BEff can be seen easily in strong lines such as \civ $\lambda 1549$, 
\ovi $\lambda 1034$, \lya $\lambda 1216$, \ciii ]$\lambda 1909$, and
\mgii $\lambda 2798$ (Figures 5a and 6). 
Taking advantage of the high signal-to-noise ratio of the composite spectra, 
we also detected this relation for weaker emission lines such as 
\heii $\lambda 1640$.
Furthermore, emission lines of the same ion but at different wavelengths 
(\oiii ]$\lambda 1663$, \ob\ and \heii $\lambda 1640$, \heii $\lambda 4686$)
possess nearly identical slopes (Tab.\,1).
For the first time, we report on the BEff for \niv ]$\lambda 1486$,
\oiii ]$\lambda 1663$, and \niii ]$\lambda 1750$.
In spite of the large scatter, the broad Fe emission features in the 
ultraviolet show some indications of a BEff, while within the errors, the
optical \feii\ emission lacks a BEff.
In addition to the broad emission lines, we detect the Baldwin Effect also
for the prominent forbidden emission line \ob\ (Figure 6), which is regarded 
as a typical narrow-line region (NLR) emission line.
In contrast to the strong BEff in most of the emission lines 
visible in Figures 5a, 5b, and 6, the Balmer lines \hgamma\ and \hbeta\ show
only weak indications of a Baldwin Effect, which might be restricted to the 
line center of the line profiles. Close inspection of the \nv $\lambda 1240$ 
emission line (Figures 5a and 8) also indicates that this line does not 
exhibit a BEff like other high ionization lines. 
Lines having the same equivalent widths in each individual spectrum will 
cancel out in difference spectra (provided that their profiles are also the
same), while lines whose equivalent widths are a function of luminosity will 
still be present in difference spectra. Note that \nv $\lambda 1240$ has on 
average cancelled out in the difference spectra as shown in Figure 8.
Hence, the difference spectra near \nv\ and the careful deblending of the 
\lya -\nv $\lambda 1240$ emission line complex both reveal that 
\nv $\lambda 1240$
remains nearly constant in $W_\lambda $ over $\sim 6$ orders of magnitude in 
continuum luminosity.

\subsection{Trends with Ionization Energy}

Zheng, Fang, \& Binette (1992), Espey et al.\,(1993), and Espey \& Andreadis 
(1999) have suggested that there is a correlation between the ionization 
energy, 
$\chi _{ion}$, of a specific emission line and the strength (slope) of the 
BEff. In contrast to Espey \& Andreadis (1999), we use the ionization energy 
necessary to create the ion for collisionally excited lines like \civ $\lambda
1549$ and \mgii $\lambda 2798$, and the ionization energy needed to ionize the
ion for recombination lines like \lya\ and \heii .

The BEff slope of the lines under study shows a relatively close linear 
correlation with $\chi _{ion}$ (Figures 7a,\,b).
We determined the slope $\beta $, where log\,$W_\lambda \propto 
\beta \cdot\,{\rm log}~ \lambda\,L_\lambda (1450{\rm \AA })$, 
of the BEff from a linear least square fit to
log\,$W_\lambda $ vs. log\,$\lambda\,L_\lambda (1450{\rm \AA })$.
The fits were calculated taking into account the full uncertainties in 
log\,$W_\lambda $.
First, we calculated the slopes across the entire luminosity range, shown by 
the dotted line in Figures 7a and 7b.
However, several of the emission lines indicate that the slope of the BEff is 
flatter at lower luminosities ($\log\,\lambda\,L_\lambda (1450 {\rm \AA }) 
\la 44$).
Therefore, we calculated additional linear fits for the luminosity range 
log\,$\lambda\,L_\lambda (1450{\rm \AA }) \ga 44$ (dashed-dotted lines).
It is interesting to note that the lines showing a change in the BEff slope
have high ionization energies with $\chi_{ion} \ga 25 - 30$\,eV. The slopes of
the log\,$W_\lambda $ vs. log\,$\lambda\,L_\lambda (1450{\rm \AA })$ 
relations are listed in Table 1 and plotted against $\chi _{ion}$ in Figure 9
for the entire range (filled symbols) and for 
log\,$\lambda\,L_\lambda (1450{\rm \AA }) \ga 44$ (open symbols).
Notice, that the slope of the BEff becomes steeper for increasing $\chi_{ion}$.
However, as has already been shown by earlier studies, \nv $\lambda 1240$ 
deviates significantly from this trend (Espey \& Andreadis 1999). 
\nv $\lambda 1240$ stays constant or shows a slightly decreasing W$_\lambda$
towards higher luminosity.

The slope $\beta $ for \ovi $\lambda 1034$ is of particular interest for 
testing the hypothesis of a steeper BEff towards higher ionization energies
since $\chi_{ion} (O^{+5})$ is more than twice $\chi_{ion} (He^{+2})$ and 
nearly $15$ times $\chi_{ion} (Mg^+)$. Assuming a constant
ratio of \lyb /\ovi $\lambda 1034 = 0.30$ does not significantly change the 
slope of the BEff for \ovi $\lambda 1034$. 
However, $\beta $(\ovi $\lambda 1034$) would become steeper by nearly $25$\,\%\
if \lyb\ follows the same BEff as \lya\ and \lyb /\lya =$0.059 \pm 0.04$ 
(Laor et al.\,1995). 

To quantify the relationship of the BEff slopes $\beta $ to $\chi _{ion}$, we
calculated linear least square fits to the points in Figure 9
excluding the outlying data point for \nv $\lambda 1240$.
The resulting slope, $\eta $, is the same with ($\eta = -0.00154 \pm 0.00024$)
or without ($\eta = -0.00150 \pm 0.00030$) the \ovi $\lambda 1034$ line 
included. This result indicates that \ovi $\lambda 1034$ does not dominate
the $\beta $ versus $\chi _{ion}$ relation.

\section{Discussion}

\subsection{Comparisons with other Studies}

In general, the slopes of the Baldwin Effect derived here are consistent with 
prior investigations. 
However, the large range in both luminosity and redshift that distinguishes 
our quasar sample reveals correlations which were missed in some other studies.
In particular, we can more clearly separate the trends with $L$ and $z$, and
our use of composite spectra across a wide luminosity range more accurately
probes the BEff slopes, e.g., for weaker lines.
In the following we provide some more detailed comparisons with earlier work.

The slopes we derived for \lya $\lambda 1216$, \civ $\lambda 1549$,
\ciii ]$\lambda 1909$, and \mgii $\lambda 2798$, are in good agreement with 
prior studies (Kinney et al.\,1990; Osmer et al.\,1994; Laor et al.\,1995; 
Green 1996; Turnshek 1997; Zamorani et al.\,1992; Zheng et al.\,1995; 
Wang et al.\,1998; Espey \& Andreadis 1999). 
The steeper slopes for \civ $\lambda 1549$ in Osmer et al.\,(1994), 
Laor et al.\,(1995), and Green (1996), $\beta = -0.23$ compared to 
$\beta =-0.13$ here, might be caused by the luminosity range of their quasar 
samples, $\lambda\,L_\lambda (1450{\rm \AA }) \ga 10^{44}$ erg\,s$^{-1}$.
Using only the $W_\lambda$(\civ) measurements for 
$\lambda\,L_\lambda (1450{\rm \AA }) \ga 10^{44}$ erg\,s$^{-1}$ in our
sample, we calculate a slope of the Baldwin Effect of 
$\beta = -0.20\pm 0.03$ for \civ $\lambda 1549$ (Tab.\,1).

A lack of the Baldwin Effect for \civ $\lambda 1549$ and \ciii ]$\lambda 1909$
is reported by Wilkes et al.\,(1999). They conclude that this is caused by the
Narrow-Line-Seyfert\,1 galaxies (NLSy1s) in their sample which decrease the 
strength of the Baldwin Effect; a significant BEff is found for both \civ\ and
\ciii ] if the NLSy1s galaxies in their sample are omitted. 
We note that these NLSy1s are also low luminosity AGN.
As can be seen in Figures 5a and 5b, the Baldwin relation is significantly 
flatter
for low luminosities than at higher luminosities. Any AGN sample favoring a 
narrow luminosity range can dilute or bias the derived BEff.
Similarly, the steeper BEff slopes reported by Laor et al.\,(1995), for 
\oi $\lambda 1305$, \cii $\lambda 1335$, \siiv $\lambda 1402$, and
\heii $\lambda 1640$ can be attributed to their smaller luminosity range
($\lambda\,L_\lambda (1350{\rm \AA }) \simeq 10^{45} - 10^{48}$ erg\,s$^{-1}$).
 
The slope for the Baldwin Effect for the high ionization emission line 
\ovi $\lambda 1034$ ($\chi _{ion} \simeq 114$\,eV) is also consistent with 
prior investigations.
The reported slope $\beta$(\ovi $\lambda 1034$) varies between $-0.30 \pm 0.04$
(Zheng et al.\,1995; Turnshek 1997) and $-0.15 \pm 0.08$ (Laor et al.\,1995), 
with most recent studies favoring a less steep anti-correlation 
($-0.18 \pm 0.03$; Espey \& Andreadis 1999). 
Wilkes et al.\,(1999) found only a marginal 
anti-correlation of $W_\lambda ($\rm \ovi $\lambda 1034)$ and continuum 
luminosity. However, they noted that this result is based on just ten objects.
Green (1996) and Green et al.\,(2001) did not detect a significant Baldwin 
Effect for \ovi $\lambda 1034$ in their analysis of the Large Bright QSO
Survey (LBQS) data set, taking all quasar spectra into account, i.e. including
 upper limit measurements as well. 
However, using only those spectra with detected \ovi $\lambda 1034$ the 
Baldwin Effect is clearly present. 
A possible reason for the missing Baldwin Effect for \ovi $\lambda 1034$ 
in their study might be the very narrow luminosity range 
($46.2 \la \log\,\lambda\,L_\lambda (2500 {\rm \AA }) \la 47.0$) covered by 
LBQS quasars at these wavelengths.
A comparison of our Figure 7a with Figure 2 in Green et al.\,(2001) shows that
we measure the same range of \ovi $\lambda 1034$ equivalent widths for this 
luminosity, $\log\,W_\lambda $(\ovi $\lambda 1034) \simeq 1.1$. 
Hence, we conclude that the non-detection of the Baldwin Effect for this 
high-ionization line in Green et al.\,(2001) is predominatly caused by their 
very small luminosity range. 

In contrast to Wills et al.\,(1999), we detect a significant Baldwin Effect
for the prominent NLR emission line \ob\ (Fig.\,7b). The slope 
$\beta $ of the anti-correlation is nearly identical to the one which we 
derived for \oiii ]$\lambda 1663$ (Tab.\,1, Fig.\,9).
Again, the lack of a BEff for \ob\ in Wills et al.\,(1999) might be
caused by the covered luminosity range of less than two orders of magnitude.
 
In agreement with prior investigations, we found no significant Baldwin Effect
for \nv $\lambda 1240$ (Osmer et al.\,1994; Laor et al.\,1995; Turnshek 1997; 
Espey \& Andreadis 1999). Instead of a decreasing equivalent width for 
increasing continuum strength, $W_\lambda ({\rm \nv }1240)$ stays nearly 
constant, even though it has a high value of $\chi _{ion} = 78$\,eV.

\subsection{Comparison with Model Predictions}

Although much observational and theoretical effort has been spent to decipher
the process which causes the Baldwin Effect, the mechanism is still unclear.
Several models have been suggested which emphasize different aspects,
i.e., contributions from optically thin clouds and luminosity-dependent 
ionization parameter, covering factor, ionizing continuum shape, and
chemical composition of the gas.

%%%%%% I %%%%%%
\subsubsection{Luminosity Dependent Ionization and Covering Factor}

The dependence of the strength of the Baldwin Effect on the ionization energy 
indicates that a decreasing covering factor towards higher luminosities is not
sufficient to explain the observed BEff for each emission line.
To explain the observed BEff for \civ $\lambda 1549$,
Mushotzky \& Ferland (1984) presented model calculations suggesting that the 
BEff is caused by a decrease of the ionization parameter, $U$, as well as 
the covering factor towards higher luminosities.
Although this model predicts a weaker Baldwin Effect for 
lower luminosity AGN, it also predicts that \lya\ will lack a BEff or show
only a weak BEff.
A luminosity dependence of the covering factor might be indicated independently
by measurements of X-ray absorption (e.g., Lawrence \& Elvis 1982). 

A variant of this picture was outlined by Shields, Ferland, \& Peterson (1995),
who suggested that the luminosity dependence could be caused by a 
luminosity-dependent covering factor of optically thin clouds which would emit
preferentially high-ionization lines. The presence of this component in the
BLR is suggested from several aspects of variability in Seyfert galaxies, as 
well as other arguments. 
While luminosity dependence of coverage by high-ionization clouds in the BLR 
is apparently still a viable scenario for understanding the BEff, this
interpretation suffers from the lack of a strong physical basis for predicting
this behavior; explanations are ad hoc or phenomenological at best.

%%%%% III %%%%%%
\subsubsection{Luminosity dependent Spectral Energy Distribution}
Several models have been advanced that focused primarily on the spectral
energy distribution (SED) of the ionizing continuum, and its consequences. 
This approach was motivated by increasing evidence that the SED of the 
continuum becomes softer for more luminous \mbox{AGN} 
(e.g., Sargent \& Malkan 1982; Binette et al.\,1989; Schulz 1992; 
 Netzer et al.\,1992; Zheng et al.\,1992; Zheng \& Malkan 1993; 
 Green 1996, 1998; Wang et al.\,1998).

Empirical models have been suggested which combine a powerlaw continuum and a 
thermal UV bump. Detailed investigations of X-ray and ultraviolet continuum
observations, i.e., the overall ionizing continuum shape, yield indications 
that the SED becomes softer for increasing luminosity. In particular, it has 
been shown that $\alpha _{ox}$, the two-point power index which connects the 
ultraviolet at 2500\,\AA\ to the X-ray continuum at 2\,keV, increases 
significantly with luminosity 
(Tananbaum et al.\,1986; Wilkes et al.\,1994; Green et al.\,1995).
This softening can be understood by a shift of the thermal UV-bump towards 
longer wavelengths in more luminous quasars, i.e., $L_{uv}$ increases more 
than $L_x$ for increasing luminosity resulting in a greater 
$\alpha _{ox}$ (cf., Binette et al.\,1989; Zheng \& Malkan 1993).

Early models of geometrically thin and optically thick accretion discs focused
on the consequences of the disc inclination and the location of gas relative 
to the disc (Netzer 1985, 1987). Although these models can not explain the
BEff for several orders of magnitude in luminosity they provide an explanation
for scatter in the $L_c$ vs. $W_\lambda $ relation at a given luminosity.
The spectral energy distribution (SED) of the accretion disc continuum depends
on the luminosity which is in turn related to the mass of the black hole and 
the accretion rate. 
These dependencies were suggested as the primary mechanism to cause the 
Baldwin Effect. 
Recently, Wandel (1999a,b) presented an evolutionary scenario which connects 
the growth of the black hole mass, the accretion rate and the continuum 
luminosity. As the mass of the black hole increases and consequently also the 
luminosity, the peak temperature of the UV bump decreases. This results in a 
shift of the peak of the thermal UV bump to longer wavelengths, i.e., the 
continuum becomes softer. Netzer et al.\,(1992) calculated the 
log\,$W_\lambda (\lya )$ vs. log\,$L_\lambda (1216 {\rm \AA })$ relation for a
wide range of black hole masses, accretion rates, and inclination angles in
geometrically thin accretion disc models.
The spread of the calculated $W_\lambda (\lya )$ at a given luminosity 
is mainly caused by the different inclination angles of the accretion disc.
Assuming that our composite spectra represent an average over a variety of 
disc inclinations, the BEff which we measured for \lya\ is consistent with 
their model predictions.

The observed ionization dependence of the Baldwin Effect provides strong 
evidence for a luminosity dependent SED of the ionizing continuum
and thus favors accretion disc models. 
However, the detailed models discussed by Wandel (1999b) show that the SEDs of
accretion discs depend in a complex way on the mass of the black hole and the 
accretion scenario.
In particular, the cutoff energy of the thermal UV-bump is expected to depend 
on the black hole mass.
These models also predict that luminous quasars with massive black holes 
should show low cutoff energies, while Seyfert\,1 galaxies with intermediate 
massive black holes should display high cutoff energies. The low luminosity 
NLSy1 galaxies with low mass black holes should also exhibit high cutoff 
energies. Different accretion scenarios would introduce additional scatter to 
this relation.

In summary, accretion disc models provide a reasonable explanation of the 
Baldwin Effect, in particular for the ionization energy $\chi _{ion}$ 
dependent strength. 
Although these models indicate that the luminosity-dependent SED is a major 
mechanism which drives the BEff, the large possible parameter space allows for
additional effects. Evidence for effects in addition to variations of the 
spectral energy distribution is given by the failure to explain the lack of a 
Baldwin Effect for another high ionization emission line, \nv $\lambda 1240$.

%%%%%% IV %%%%%%
\subsubsection{The Influence of Metallicity Variations}

Baldwin et al.\,(1995) and Korista et al.\,(1997, 1998) introduced the 
``locally optimally-emitting cloud'' (LOC) model, which assumes that the BLR 
cloud ensemble covers a wide range of internal densities and occurs over a wide range in
distance from the central continuum source. 
Under these conditions the emitted spectrum is controlled by powerful 
selection effects and the typical observed quasar spectrum can be naturally 
produced (Korista et al.\,1997, 1998). 

Korista et al.\,(1998) computed for a wide range of different continuum
slopes ($F_\nu \propto \nu ^\alpha, -2 < \alpha < -1$), densities ($n_e = 10^8 
\,\,{\rm to}\,\,10^{12}$\,cm$^{-3}$), and metallicities (0.2 to 10 times solar)
the resulting emission line spectra. 
In recent years there has been growing evidence that the chemical composition
of quasar gas can reach several times solar metallicity with a trend of 
higher metallicities in more luminous quasars (Hamann \& Ferland 1993, 1999; 
Ferland et al.\,1996, Dietrich et al.\,1999; Dietrich \& Wilhelm-Erkens 2000).
The emission lines studied by Korista et al.\,(1998) follow the general trend 
for $W_\lambda $ diminishing with softer continuum shapes. Assuming the
softer continua are related to higher luminosities, this yields the Baldwin
Effect.
However, in contrast to those emission lines, \nv $\lambda 1240$ lacks a 
Baldwin Effect even though its high ionization energy $\chi _{ion} = 78$\,eV 
suggests that the decrease of $W_\lambda $(\nv ) should be strong.
Korista et al.\,(1998) introduced metallicity as an additional parameter to
provide a possible explanation for the lack of a Baldwin Effect in 
\nv $\lambda 1240$.
They found in their calculations that the equivalent width of \lya , 
\civ $\lambda 1549$, or \ovi $\lambda 1034$ show only a weak dependence on 
the gas metallicity, but \nv $\lambda 1240$ strongly depends on it. 

In particular, they assume that nitrogen scales like a secondary element, so 
that its abundance relative to the other metals increases linearly with the 
overall metallicity (e.g., N/O $\propto$ O/H, see Hamann et al.\,2002).
Korista et al.\,(1998) assume that the metallicity increases with increasing
AGN luminosity (Hamann \& Ferland 1999). With this abundance behavior,
Korista et al.\,(1998) showed that the equivalent width of \nv $\lambda 1240$ 
is not expected to display a BEff. 
Therefore, the decreasing equivalent width that should 
occur in \nv $\lambda 1240$ as part of the normal BEff is approximately 
compensated by the increasing relative abundance of nitrogen. 
The net result is that \nv $\lambda 1240$ shows essentially no BEff. 

Overall, the BEff slopes that we derived for the emission lines in our quasar 
sample are in good agreement with the predictions by Korista et al. (Table 1). 
The observed slopes of \civ $\lambda 1549 $ and \heii $\lambda 1640$ are 
smaller than expected, while the slope in \niii ]$\lambda 1750$ is steeper. 
However, if only measurements with 
$\lambda L_\lambda (1450 {\rm \AA }) \ga 10^{44}$ erg\,s$^{-1}$ are taken into
account, the observations yield a
comparable steep anti-correlation as predicted by Korista et al.\,(1998).
We will 
discuss these comparisons further in a forthcoming paper on the trends in 
metallicity in this dataset (see also \S6.4 below).

%%%%%%%%
\subsection{Ionization Dependence and Curvature of the Baldwin Effect}

A common prediction of models that involve a softer continuum SED towards 
higher luminosities is a steeper slope of the Baldwin Effect for lines with 
increasing ionization energy $\chi _{ion}$ (Zheng et al.\,1992; 
Zheng \& Malkan 1993; Korista et al.\,1998; Wandel 1999a). 
Figures 7a,b and 9 provide clear evidence for this prediction.
High ionization lines like \civ $\lambda 1549$, \heii $\lambda 1640$, and 
\heii $\lambda 4686$, and especially \ovi $\lambda 1034$ are particularly 
valuable.
These emission lines respond to photoionizing energies of $\sim 48$\,eV, 
$\sim 54$\,eV, and $\sim 114$\,eV, respectively. Hence, they can be used to
probe the location of the cutoff energy of the thermal UV-bump. 
The strong BEff in these high-ionization lines is expected if the ionizing
continuum becomes softer with increasing continuum luminosity. In this case
the relative number of ionizing photons with  $h\,\nu \ga 50$\,eV decreases 
as the UV bump is moving towards longer wavelengths.
While there is only a marginal trend for a slightly steeper Baldwin Effect 
for \heii\ than for \civ , the higher ionization \ovi $\lambda 1034$ line
displays a significantly stronger BEff (Fig.\,9).
The decrease of the relative number of ionizing photons with a softer 
continuum, caused by the shift of the UV bump to longer wavelengths, results 
in weaker high-ionization lines compared to lower ionization ones.
The shift of the UV bump to longer wavelengths will affect emission lines at 
higher ionization energy first. In addition, the relative continuum strength 
beneath the emission lines will be enhanced, starting at short ultraviolet
wavelengths. Both effects together result in smaller $W_\lambda $ for higher 
luminosities.

In addition to the general trend of steeper BEff for higher $\chi _{ion}$, 
the curvature or flattening of the BEff towards lower luminosities can be 
accommodated in the framework of some of the luminosity-dependent SED models 
(e.g., Netzer et al. 1992; Wandel 1999a).
The flattening at low luminosities has led to suggestions that Seyfert\,1 
nuclei do not participate in the Baldwin Effect but have instead equivalent 
widths independent of luminosity (e.g., Wampler et al.\,1984).
A close inspection of Figures 7a,b provides some evidence for the tendency
of a flatter slope $\beta $ of the BEff at low luminosities in comparison to
higher luminosities, as can be seen for several emission lines
(\siiv $\lambda 1402$, \niv ]$\lambda 1486$, \civ $\lambda 1549$,
 \heii $\lambda 1640$, \oiii ]$\lambda 1663$, \niii ]$\lambda 1750$,
 \heii $\lambda 4686$).
The turnover appears to occur near
$\lambda \log\,L_\lambda (1450 {\rm \AA }) \simeq 44$, in quite good agreement
with predictions (Wandel 1999a).
Within this context it is interesting to note that the steepening of the 
Baldwin Effect for $\log\,\lambda\,L_\lambda (1450 {\rm \AA }) \simeq 44$,
is observed for emission lines with $\chi _{ion} \ga 25 - 30$\,eV.

%%%%%%
\subsection{Individual emission line pairs}
For several elements we have measured more than one emission line strength.
The comparison of the strength of the Baldwin Effect may provide information 
about the emission line region and the mechanism which causes the 
observed anti-correlation of $W_\lambda $ and continuum luminosity.

For hydrogen we have measured two Balmer emission lines (H$\gamma $, H$\beta $)
and \lya . The slope of the Baldwin Effect of H$\gamma $ and H$\beta $ is
very similar and within the uncertainties consistent with zero, i.e. no 
Baldwin Effect for Balmer lines. The missing Baldwin Effect for these
lines provide some argument against a luminosity dependent covering factor,
at least for the gas component which is the dominant contributor to these 
emission lines.
While the equivalent width $W_\lambda$(H$\beta$) remains nearly constant, the 
flux ratio Ly$\alpha$/H$\beta$ decreases by a factor $\sim 1.5$ from low to 
high luminosities ($\sim 8.5$ to $\sim 5.5$). The different behaviour of the 
Balmer lines and Ly$\alpha$ might be also related to the complex physical 
processes of Ly$\alpha$ and H$\beta $ emission (Netzer et al.\,1995). The 
prominent BEff of Ly$\alpha$ and the lack of a BEff for H$\beta$ may be caused
in part by a relative stronger increase of the local continuum beneath 
Ly$\alpha $ compared to the optical wavelength range of H$\beta$.
However, \lya\ shows a significant BEff which fits well in the slope $\beta $ 
vs. $\chi _{ion}$ relation (Fig.\,7a,\,9).
\lya\ might be blended with \heii $\lambda 1216$ and \ov ]$\lambda 1218$ 
(Ferland et al.\,1992; Shields et al.\,1995). 
For standard BLR conditions the contribution of \heii $\lambda 1216$ and 
\ov ]$\lambda 1218$ are negligible (less than $\sim 10$\,\% , 
priv.comm.\,K.\,Korista).
But for higher densities and higher ionization parameter U, particularly for 
optically thin conditions, these emission lines can be quite strong compared 
to \lya\ (Shields et al.\,1995).
Because \ov ]$\lambda 1218$ has an ionization energy of 
$\chi_{ion} = 77.4$\,eV, this emission line might show a strong BEff.
If optically thin gas is a significant fraction of the BLR gas, it might be 
possible that the moderate BEff of \lya\ might be caused at least in part
by \ov ]$\lambda 1218$.
However, if \ov ]$\lambda 1218$ is a significant contamination of \lya\ it is 
expected from the results in Shields et al.\,(1995) that 
\heii $\lambda 1640$ and \civ $\lambda 1549$ should be stronger than observed. 
In addition, variability behavior in some objects suggests that the optically 
thin gas may contribute most strongly to the line wings (e.g., Ferland et 
al.\,1990; Peterson et al.\,1993), but the part of the emission lines that 
changes the most with luminosity has a narrow FWHM (e.g., Osmer et al.\,1994).
Although we cannot exclude that \ov ]$\lambda 1218$ provides some contribution
to \lya , it does appear that \lya\ itself shows a moderate Baldwin Effect in 
contrast to the Balmer emission lines.

The two emission lines of He$^+$  which we measured, \heii $\lambda 1640$ and 
\heii $\lambda 4686$, show very similar slopes for the Baldwin Effect. 
This behavior is expected for \heii $\lambda 1640$ and \heii $\lambda 4686$
as typical recombination lines.
The similar slope of the BEff provides also some evidence that
luminosity-dependent radiative transfer effects, collisional effects or dust 
effects can not affect these emission lines.

Another line pair we measured is \oiii ]$\lambda 1663$ and
[\oiii ]$\lambda 5007$. While the ionization energy necessary to create 
O$^{+2}$ is $\chi _{ion} = 35.1$\,eV, the excitation energies of these lines
differ by a factor of $\sim 3$.
If the ionizing continuum becomes softer for increasing luminosity, the 
temperature in the emission line region should drop and hence 
\oiii ]$\lambda 1663$ should be more affected than [\oiii ]$\lambda 5007$. 
Therefore, we might expect \oiii ]$\lambda 1663$ to have a steeper BEff
than [\oiii ]$\lambda 5007$. Observationally their BEffs are the same.
However, the comparison of these two lines is very complicated because they
differ in critical density by several orders of magnitudes 
($n_e^{crit}\simeq 7\times 10^5$\,cm$^{-3}$ for [\oiii ]$\lambda 5007$ and
 $n_e^{crit}\simeq 3\times 10^{10}$\,cm$^{-3}$ for \oiii ]$\lambda 1663$). 
In particular, [\oiii ]$\lambda 5007$ forms in the narrow line region
(NLR), which might be heated and ionized additionally by shocks, and is, in 
any case, spatially distinct from the BLR.
The different kinematics of the \oiii\ emitting gas is also given by the 
significantly different profiles we used to measure the emission line flux and 
equivalent width. The \oiii ]$\lambda 1663$ emission line was fit with a broad
and narrow Gaussian profile (with an average FWHM of 
$5140\pm460$\,km\,s$^{-1}$ and $1865\pm290$\,km\,s$^{-1}$, respectively). The 
two Gaussian components we used to fit [\oiii ]$\lambda 5007$ result in a 
profile different compared to \oiii ]$\lambda 1663$, with average FWHMs of
$1490\pm50$\,km\,s$^{-1}$ and $480\pm25$\,km\,s$^{-1}$.
The overall average FWHM of the fitted \oiii ]$\lambda 1663$ and 
[\oiii ]$\lambda 5007$ lines are $4130\pm450$ km\,s$^{-1}$ and 
$725\pm100$ km\,s$^{-1}$, respectively.

We have studied three emission lines of nitrogen. \nv $\lambda 1240$ lacks 
within the errors a BEff. In contrast, the intercombination lines 
\niv ]$\lambda 1486$ and \niii ]$\lambda 1750$ exhibit a Baldwin Effect that 
is consistent with the overall trend of $\beta $ vs. $\chi _{ion}$ (Fig.\,9).
It might be possible that additional effects influence the emission properties
of these nitrogen intercombination lines at higher luminosities. 
We will discuss these issues in our forthcoming paper on elemental abundances.

%%%%%
\section{Conclusion}

We have investigated a large sample of $744$ type\,1 AGN covering the redshift 
range from $0 \leq z \leq 5$ and nearly 6 orders of magnitude in continuum 
luminosity. To enhance the signal-to-noise ratio, minimize the influence of 
peculiarities of individual quasars, and investigate weak as well as strong
emission lines, we computed composite spectra representing narrow intervals in
redshift and luminosity.
The emission line fluxes were derived using multi-component Gaussian fits  
after removing a powerlaw continuum fit, a Balmer continuum emission template,
and a UV and optical iron emission template from the composite spectra. 
Our main results are the following.

\begin{itemize}
\item In composite spectra spanning the full redshift range at nearly constant 
      luminosity we detect no strong trend in the line $W_\lambda $ with 
      redshift i.e., with cosmic time.
      However, there is a marginal tendency for the highest redshift quasars 
      ($z \ga 4$) to show slightly stronger emission lines than their 
      counterparts at lower redshift.
\item In the composite spectra ranked by luminosity we find a significant 
      Baldwin Effect in nearly all emission lines in the ultraviolet to 
      optical domain. The only exceptions are \nv $\lambda 1240$, H$\beta $, 
      H$\gamma $, and optical \feii\ which remain constant in $W_\lambda $ 
      within the uncertanties.
      The lack of a BEff for the high-ionization feature NV$\lambda 1240$ 
      suggests that the chemical composition of the gas is an additional 
      parameter that can strongly influence the equivalent width of this and 
      possibly other lines.
\item We detect a strong Baldwin Effect for the prominent NLR emission line
      [\oiii ]$\lambda 5007$. The strength of the BEff for [\oiii ]$\lambda
      5007$ is very similar to the BEff measured in \oiii ]$\lambda 1663$.
\item The slope $\beta $ of the Baldwin Effect, where 
      log\,$W_\lambda \propto \beta \cdot$\,log\,$\lambda\,L_\lambda $,
      shows a significant correlation 
      with the ionization energy, $\chi _{ion}$, needed to produce the lines.
\item The slope of the Baldwin Effect, $\beta $, tends to be steeper at higher
      luminosities, 
      $\lambda\,L_\lambda (1450 {\rm \AA }) \ga 10^{44}$\,erg\,s$^{-1}$,
      compared to the lower luminosity regime. 
\item The Baldwin Effect, its steepening towards higher luminosities, and the 
      correlation of the slope $\beta $ with $\chi_{ion}$ can all be well 
      explained in the context of a luminosity 
      dependent spectral energy distribution of the ionizing continuum.
      Assuming that the SED can be described as
      a combination of a powerlaw continuum and a thermal UV bump, the ionizing
      continuum becomes softer for increasing luminosity as the UV bump is 
      shifted to longer wavelengths. 
      This behaviour can be explained with accretion disc models as suggested
      by Netzer et al.\,(1992) and Wandel (1999a,b).
\end{itemize}

%% The \notetoeditor{TEXT} command allows the author to communicate
%% information to the copy editor.  This information will appear as a
%% footnote on the printed copy for the manuscript style file.  Nothing will
%% appear on the printed copy if the preprint or
%% preprint2 style files are used.

\acknowledgments

We are grateful to numerous colleagues for providing many of these spectra in 
digital form, in particular
J.A.\,Baldwin, T.A.\,Boroson,
M.R.\,Corbin,
A.\,Laor,
W.L.W.\,Sargent, D.P.\,Schneider, C.C.\,Steidel,
R.J.\,Weymann, and
W.\,Zheng.
We also thank Desika Narayanan for reducing the spectra obtained 
at Lick Observatory.
MD and FH have benef\/ited from support from NASA grant NAG 5-3234 and
NSF through AST-99-84040.
MV greatfully acknowledges financial support from the Columbus Fellowship.

\clearpage

Fig.1 ---
Redshift distribution of the AGN sample as a function of intrinsic luminosity 
at $\lambda = 1450$\,\AA. 
The open symbols represent radio-loud quasars and the filled symbols 
radio-quiet quasars. The dashed horizontal (vertical) lines indicate the 
luminosity (redshift) ranges which were used to calculate composite spectra. 
At the left side (top) of the figure is the number of 
the individual spectra contributing to each composite spectrum. 
For the luminosity range 
$46.16\leq \log \lambda\,L _\lambda (1450) \leq 47.16$ composite spectra were
calculated for the redshift intervals marked by the vertical
dashed lines.

Fig.2 ---
Composite spectrum of the entire AGN sample (top panel). 
The dashed line shows a power law fit to the continuum
($1200 {\rm \AA }\leq \lambda \leq 5800$\AA )
yielding $\alpha = -0.43$ $(F_\nu \propto \nu^\alpha)$.
The bottom panel displays the number of AGN contributing to each 
wavelength element.

Fig.\,3 --- (a) Normalized composite spectra are shown for redshift bins of 
$\Delta z = 0.5$, starting at $z = 0.5$ (cf., Fig.\,1) and nearly 
constant luminosity ($46.16\leq \log \lambda\,L _\lambda (1450) \leq 47.16$).
The spectra were normalized by a power law continuum fit. 
The horizontal dashed lines indicate the continuum level for the
individual normalized composite spectra which were vertically 
shifted for better display. The normalized continuum strength is shown 
for the spectrum at the  bottom of the figure. This same scale applies
to all other spectra.

(b) Same as Fig.\,3a, but with an expanded vertical scale to display the 
dependence of the relative strength of weaker emission lines as a 
function of redshift.
Strong emission lines with flat tops are truncated for easier display.

Fig.\,4  --- (a) Line equivalent widths, $W_\lambda$, vs. redshift for nearly 
constant intrinsic continuum luminosity $\lambda L_\lambda (1450{\rm \AA })$.

(b) Same as Fig.\,4a, but for additional emission lines.

Fig.\,5 --- (a) Normalized composite spectra are shown for the luminosity bins 
($\Delta \log \lambda\,L_\lambda (1450{\rm \AA }) = 0.5\,{\rm dex}$) as 
displayed in Fig.\,1. The spectra were normalized with the corresponding power
law continuum fit. The horizontal dashed lines indicate the continuum-level 
for the individual normalized composite spectra which were vertically shifted 
for better display. The normalized continuum strength is shown for the 
spectrum at the bottom of the figure and it applies for the other spectra too. 

(b) Same as Fig.\,5a, but with an expanded vertical scale to display the 
dependence of the relative strength of weaker emission lines as a function of 
luminosity.
Strong emission lines with flat tops are truncated for easier display.

Fig.\,6 ---
Same as Fig.\,5a, but the normalized composite spectra are shown at
larger wavelengths.

Fig.\,7 --- (a) Line equivalent widths, $W_\lambda$, as a function of 
increasing continuum luminosity $\lambda\,L_\lambda (1450{\rm \AA })$.
We calculated linear fits to $W_\lambda $(L) for the entire luminosity 
range (dashed line), as well as a luminosities log\,$\lambda\,L_\lambda (1450{\rm \AA }) \geq 44$ (dashed-dotted line).

(b) Same as Fig.\,7a, but for additional emission lines.

Fig.\,8 --- Difference spectra with respect to the composite spectrum
            log\,$\lambda\,L_\lambda (1450{\rm \AA}) = 47.32$ erg\,s$^{-1}$
            are displayed
            to illustrate that $W_\lambda$(\nv $\lambda 1240$) is nearly
            constant, regardless how $W_\lambda $(\nv) is measured,
            while the other emission lines display a prominant BEff.
            The dotted line indicates the location of \nv $\lambda 1240$.

Fig.\,9 --- The slope of the Baldwin Effect as displayed in Fig.\,7a and 7b as
a function of the ionization energy $\chi _{ion}$ needed to create the 
specific ions. Different lines of the same ion (\heii $\lambda 1640$, 
\heii $\lambda 4686$ and \oiii ]$\lambda 1663$, \ob ) show nearly identical 
slopes. The filled symbols represent the slopes based on the entire luminosity
range, corresponding to the dashed lines in Figures 7a,b. The slopes of the 
Baldwin Effect for higher luminosities only
(log\,$\lambda\,L_\lambda (1450{\rm \AA }) \ga 44$) are plotted as open 
symbols.

\clearpage


\begin{figure}
%\plotone{beffpapf01.ps}
\plotone{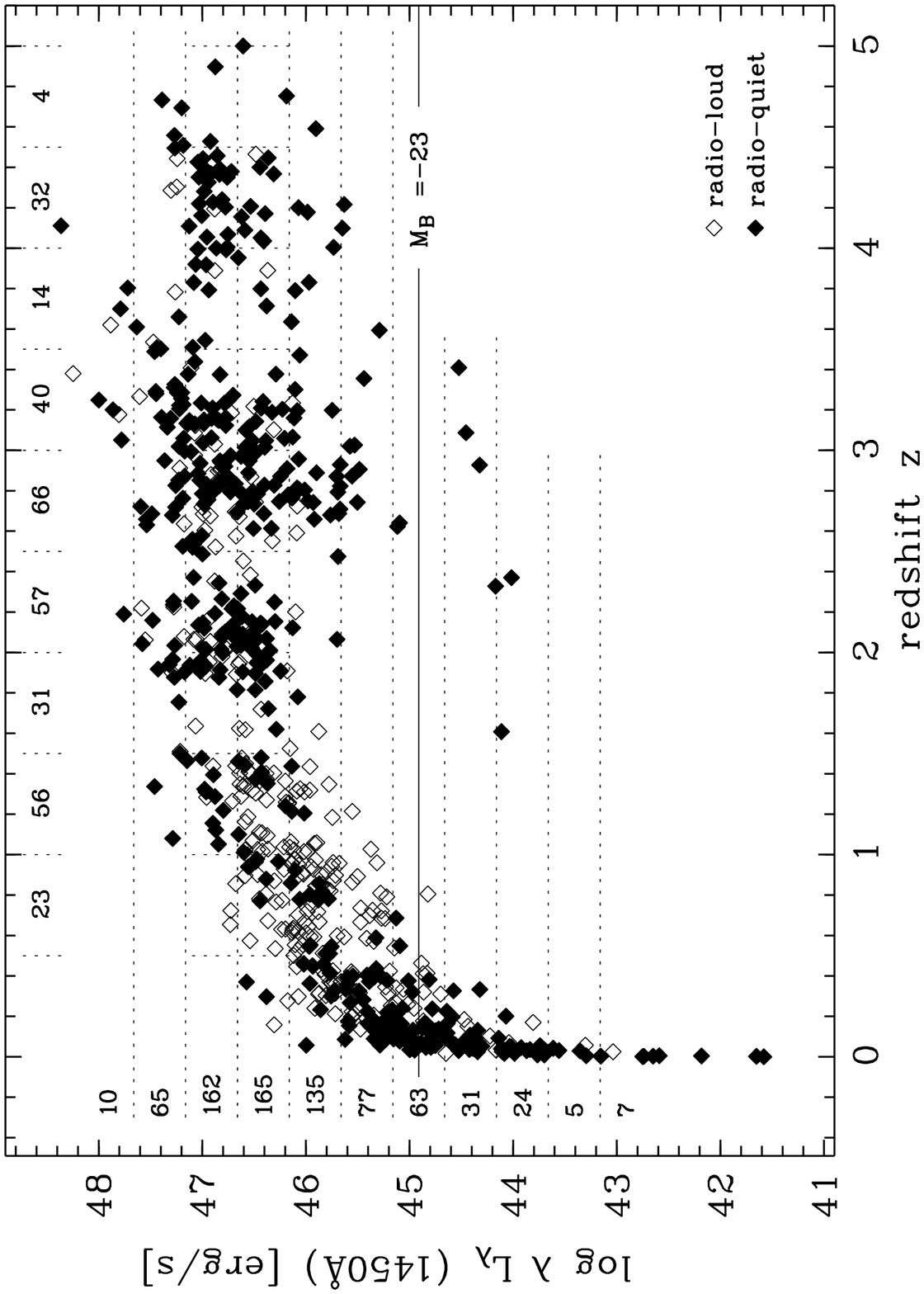}
\caption{Redshift distribution of the quasar sample as a function of 
         intrinsic luminosity at $\lambda = 1450$\,AA. 
         The open symbols represent radio-loud quasars and the filled 
         symbols radio-quiet quasars. The dashed horizontal lines indicate 
         the luminosity ranges which were used to calculate 
         composite spectra. At the left side of the figure is the number of 
         the individual spectra contributing to each composite spectrum. 
         For the luminosity range 
         $46.1\leq \log \lambda\,L_\lambda (1450) \leq 47.1$ composite spectra
         were  calculated for the redshift intervals marked by the vertical
         dashed lines.
         The number of spectra contained in each redshift bin is given at 
         the top of the figure.
\label{fig1}}
\end{figure}

\begin{figure}
%\plotone{beffpapf02.ps}
\plotone{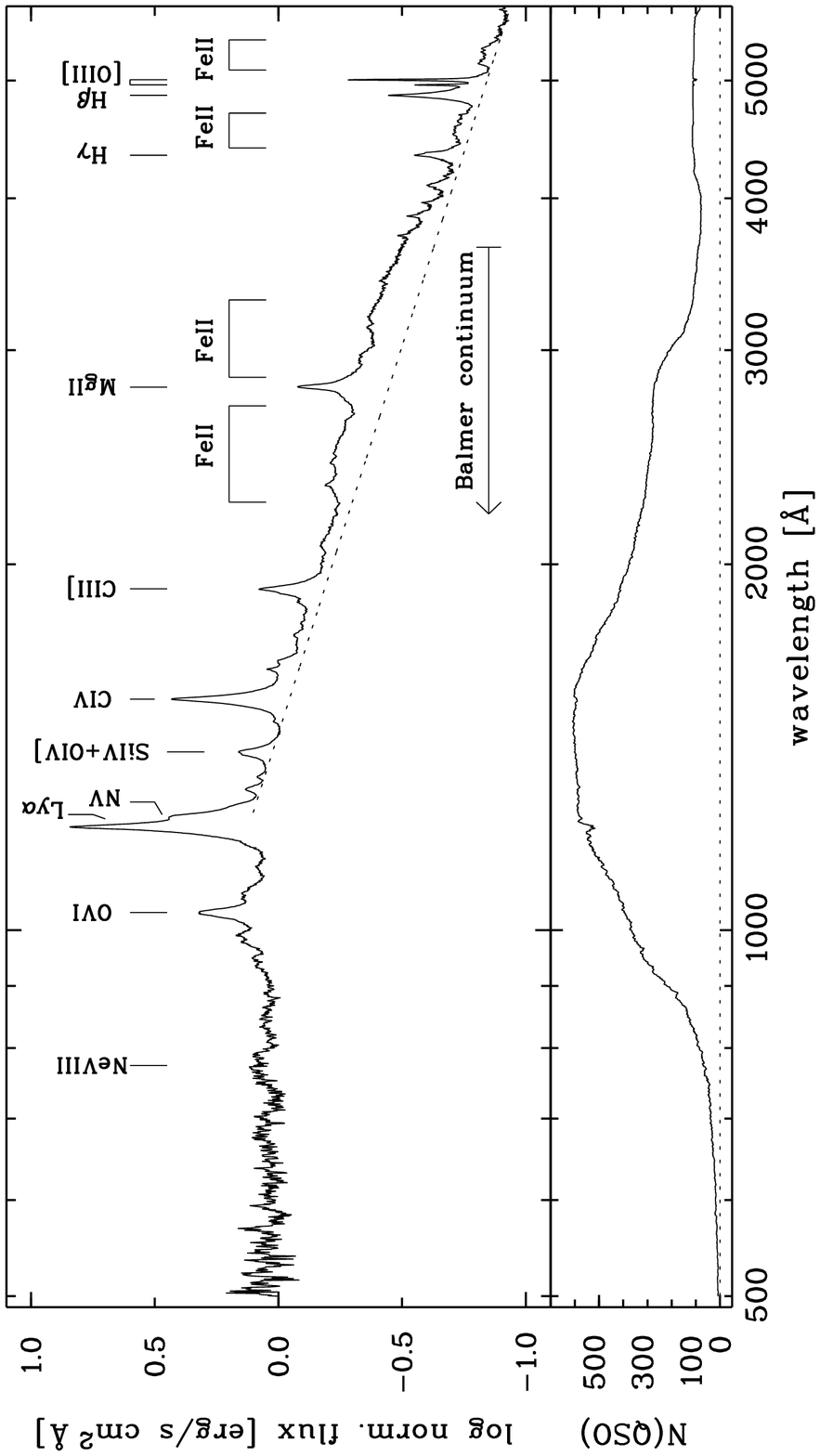}
\caption{Composite spectrum of the quasar sample (top panel). 
         The dashed line show a power law fit to the continuum (
         $1200{\rm \AA } \leq \lambda \leq 5800$\AA ) yielding 
         $\alpha = -0.43$ $(F_\nu \propto \nu^\alpha)$.
         The bottom panel displays the number of quasars contributing to each 
         wavelength element.
\label{fig2}}
\end{figure}

\begin{figure}
%\plotone{beffpapf03a.ps}{beffpapf03b.ps}
\plottwo{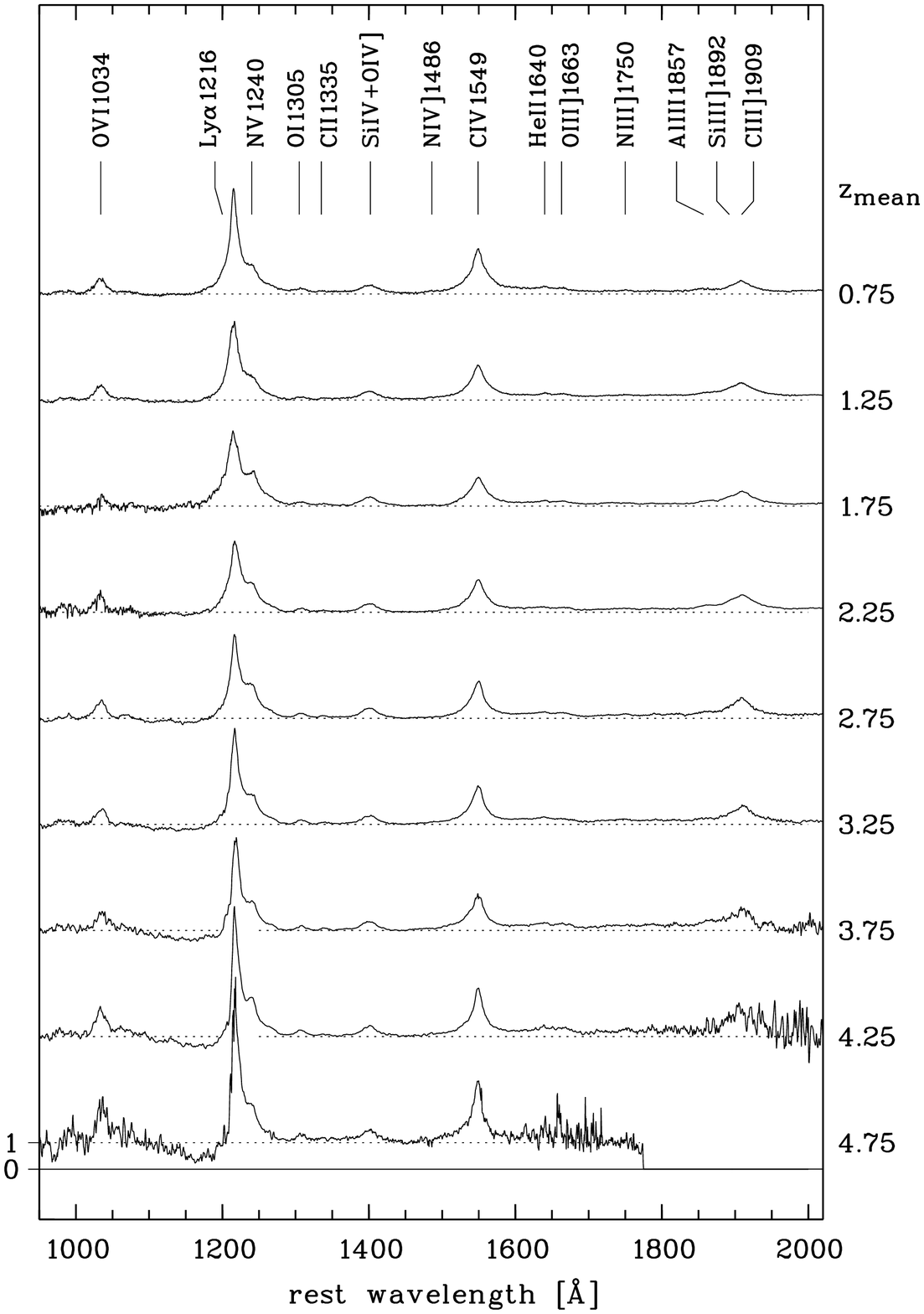}{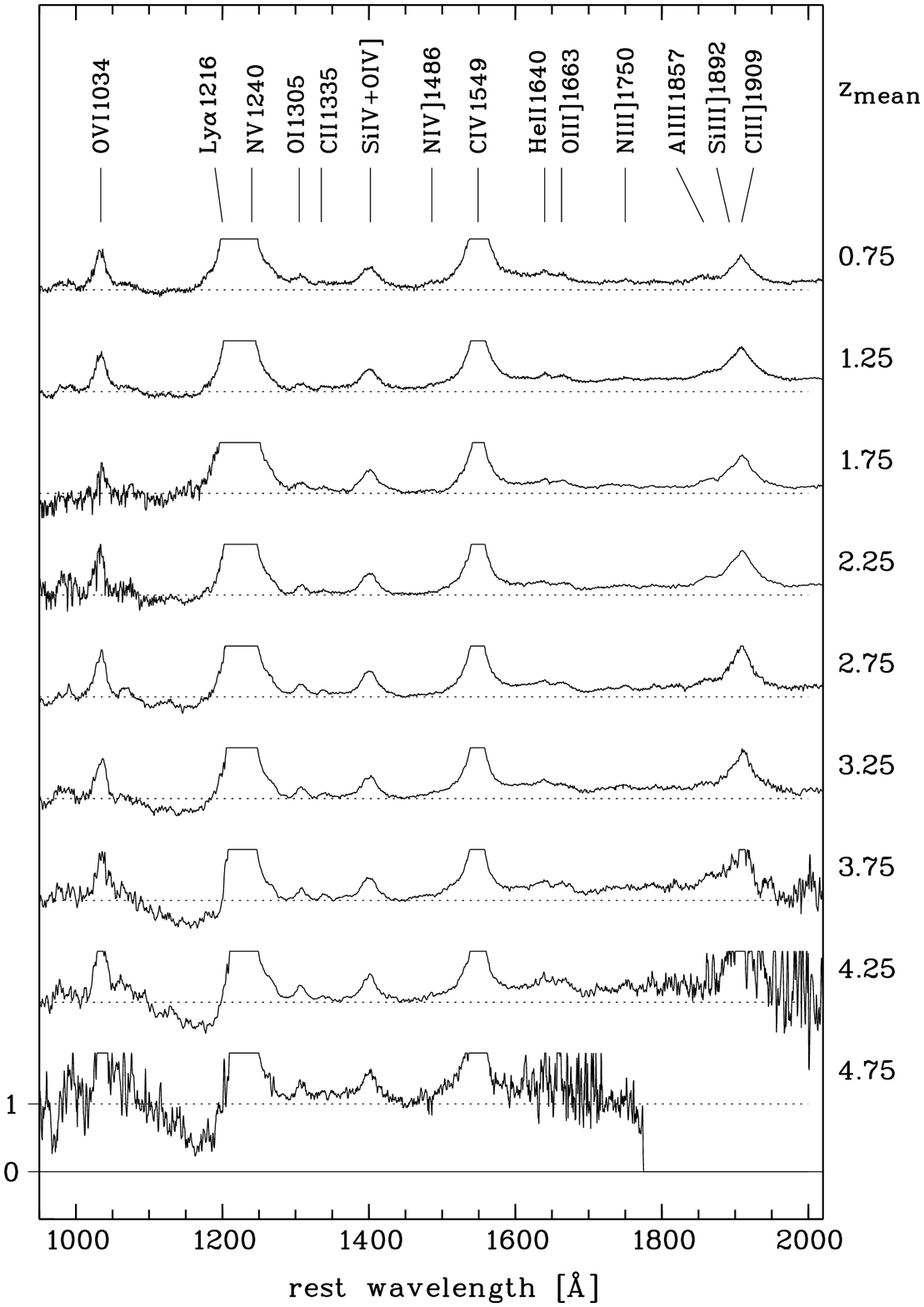}
\caption{(a) Normalized composite spectra are shown for redshift bins of 
         $\Delta z = 0.5$, starting at $z = 0.5$ (cf., Fig.\,1) and nearly 
         constant luminosity 
         ($46.1\leq \log \lambda\,L_\lambda (1450) \leq 47.1$).
         The spectra were normalized with the corresponding power law 
         continuum fit. 
         The horizontal dashed lines indicate the continuum-level for the
         individual normalized composite spectra which were vertically 
         shifted for better display.
         The normalized continuum strength is shown for the spectrum at the
         bottom of the figure and it applies for the other spectra, too. 
         (b) Same as (a), but with an expanded vertical scale to display the 
         dependence of the relative strength of weaker emission lines as a 
         function of redshift.
         Strong emission lines with flat tops are truncated for easier display.
\label{fig3}}
\end{figure}

%\begin{figure}
%%\plotone{beffpapf03b.ps}
%\plotone{f3b.eps}
%\caption{Same as Fig.\,3a, but with an expanded vertical scale to display the 
%         dependence of the relative strength of weaker emission lines as a 
%         function of redshift.
%        Strong emission lines with flat tops are truncated for easier display.
%\label{fig3b}}
%\end{figure}

\begin{figure}
%\plotone{beffpapf04af.ps}{beffpapf04bf.ps}
\plottwo{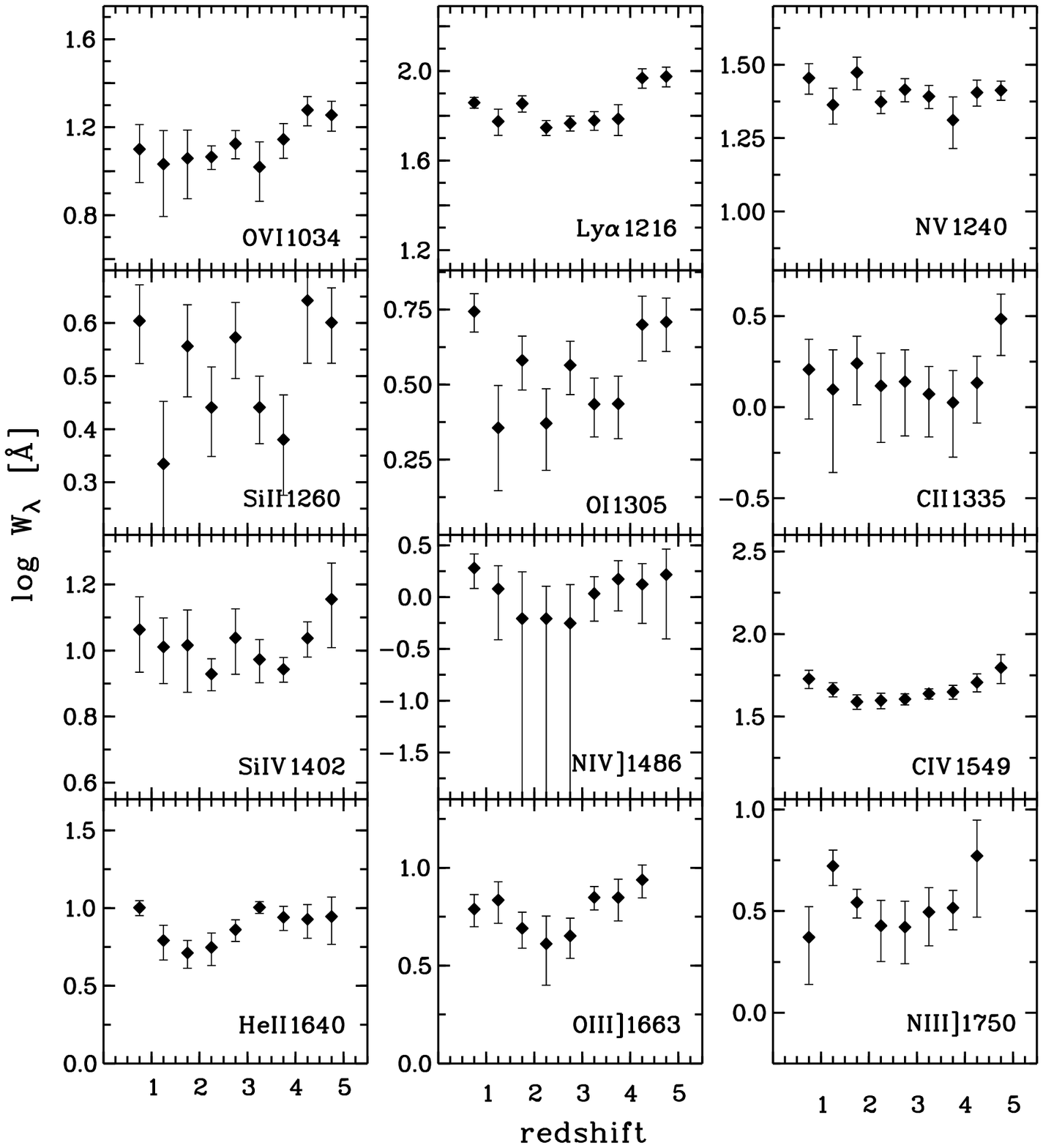}{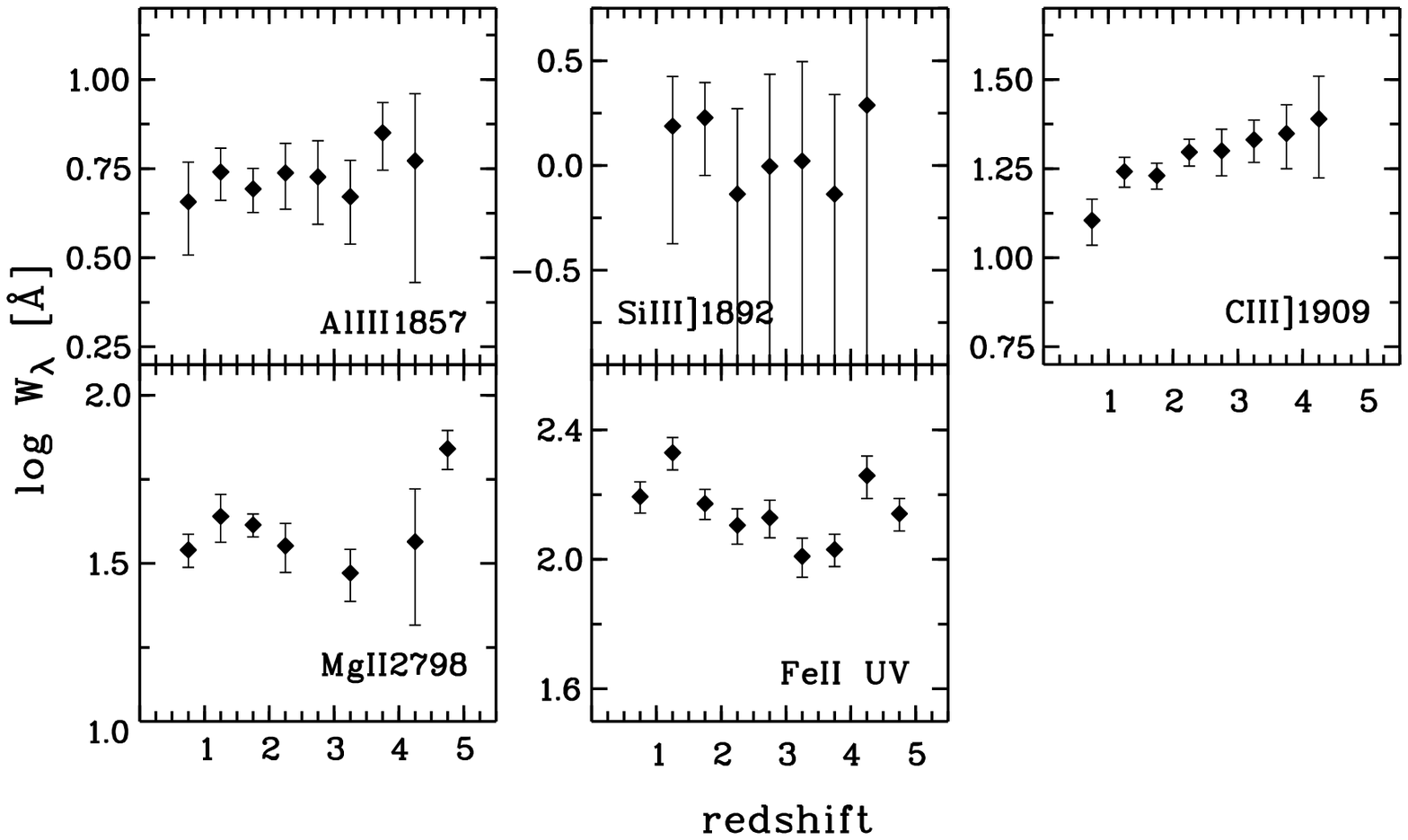}
\caption{Line equivalent widths, $W_\lambda$, vs. redshift for nearly constant
         intrinsic continuum luminosity $\lambda\,L_\lambda (1450{\rm \AA })$.
\label{fig4}}
\end{figure}

%\begin{figure}
%%\plotone{beffpapf04bf.ps}
%\plotone{f4b.eps}
%\caption{Same as Fig.\,4a.
%\label{fig4b}}
%\end{figure}

\begin{figure}
%\plotone{beffpapf05af.ps}{beffpapf05bf.ps}
\plottwo{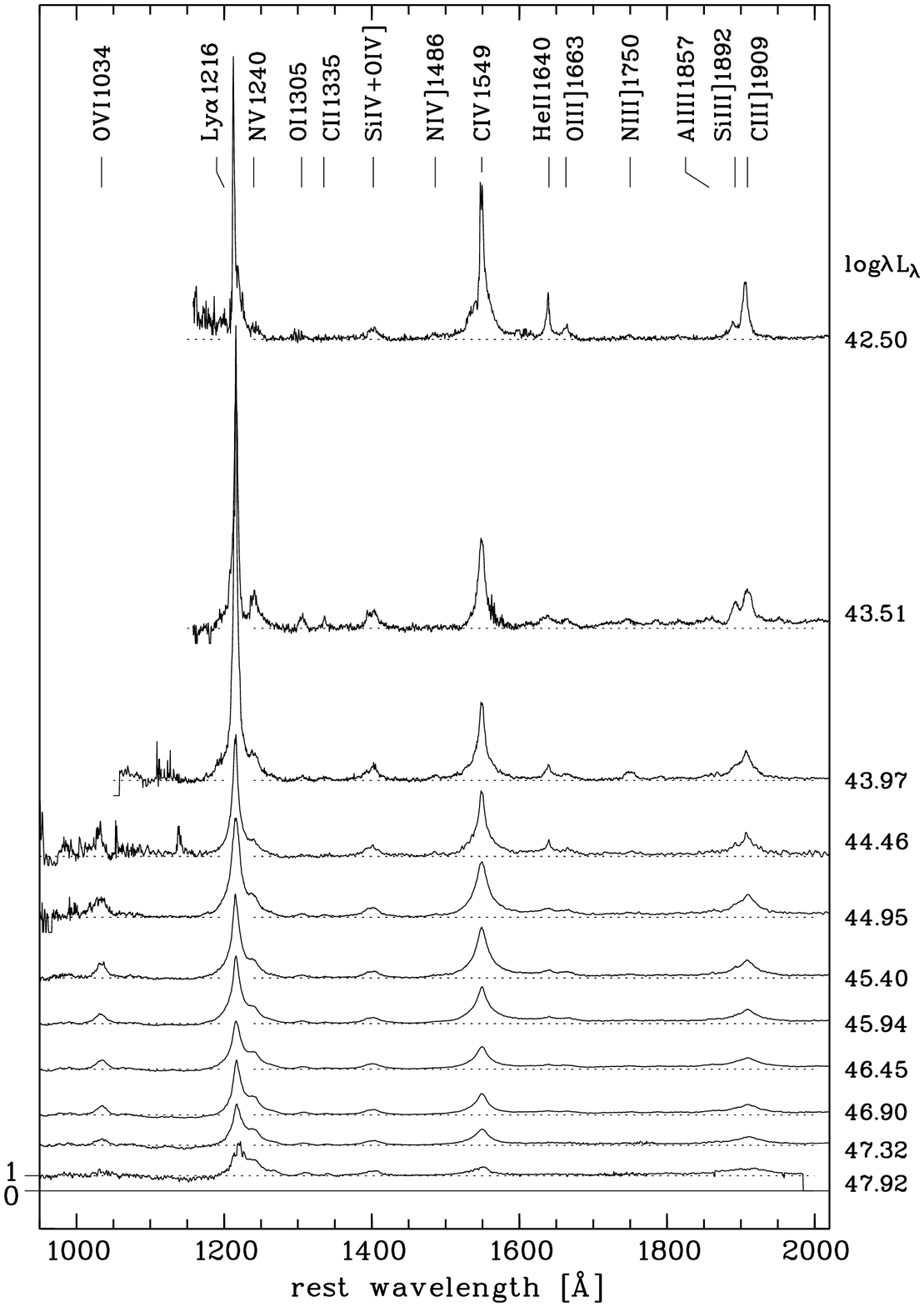}{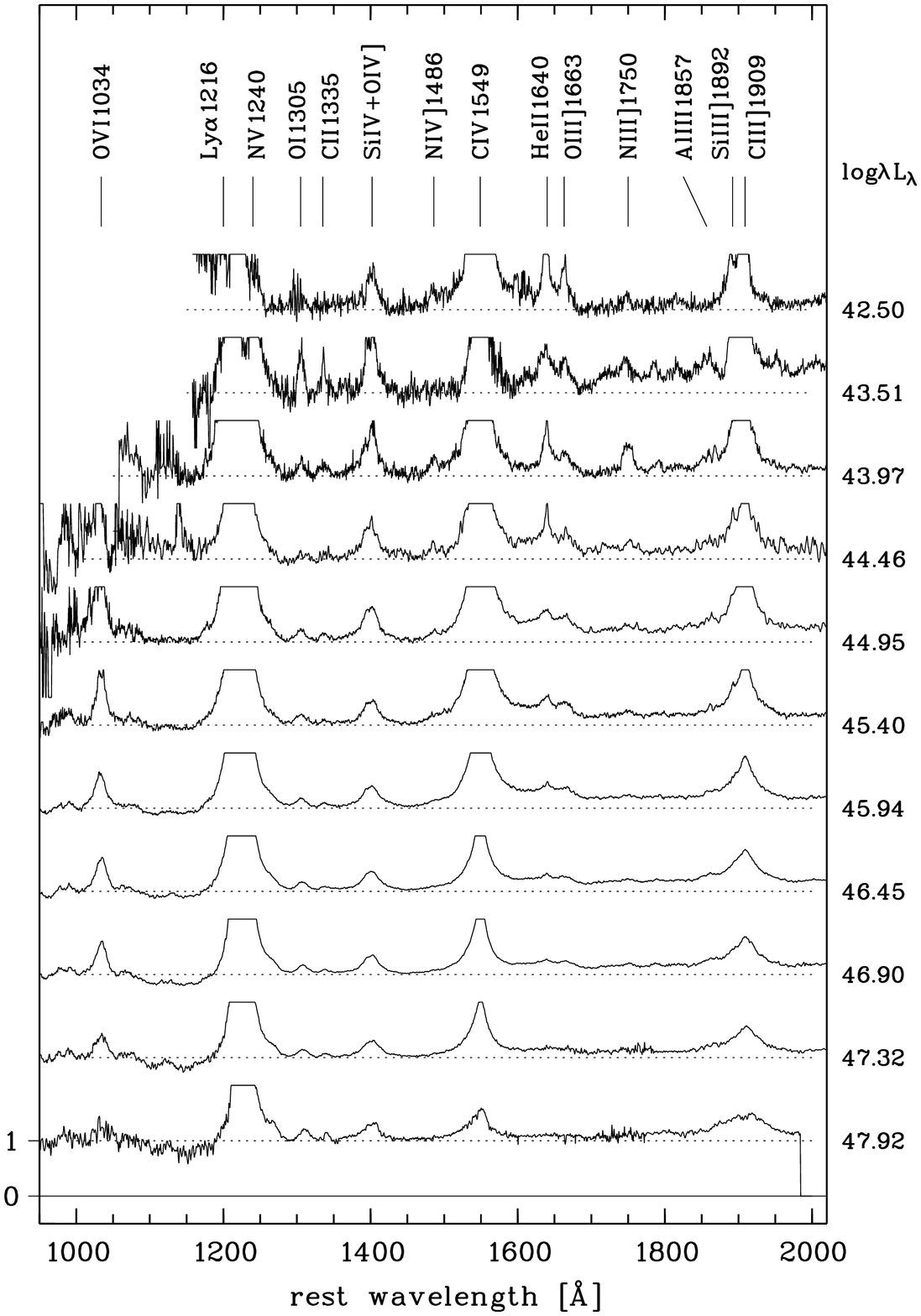}
\caption{(a) Normalized composite spectra are shown for the luminosity bins 
         ($\Delta \log \lambda\,L_\lambda (1450{\rm \AA }) = 0.5\,dex$) as 
         displayed in Fig.\,1.
         The spectra were normalized with the corresponding power law 
         continuum fit. 
         The horizontal dashed lines indicate the continuum-level for the
         individual normalized composite spectra which were vertically 
         shifted for better display.
         The normalized continuum strength is shown for the spectrum at the
         bottom of the figure and it applies for the other spectra, too. 
         (b) Same as (a), but with an expanded vertical scale to display the 
         dependence of the relative strength of weaker emission lines as a 
         function of luminosity.
         Strong emission lines with flat tops are truncated for easier display.
\label{fig5}}
\end{figure}

%\begin{figure}
%%\plotone{beffpapf05bf.ps}
%\plotone{f5b.eps}
%caption{Same as Fig.\,5a, but with an expanded vertical scale to display the 
%        dependence of the relative strength of weaker emission lines as a 
%        function of luminosity.
%        Strong emission lines with flat tops are truncated for easier display.
%\label{fig5b}}
%\end{figure}

\clearpage

\begin{figure}
%\plotone{beffpapf06f.ps}
\plotone{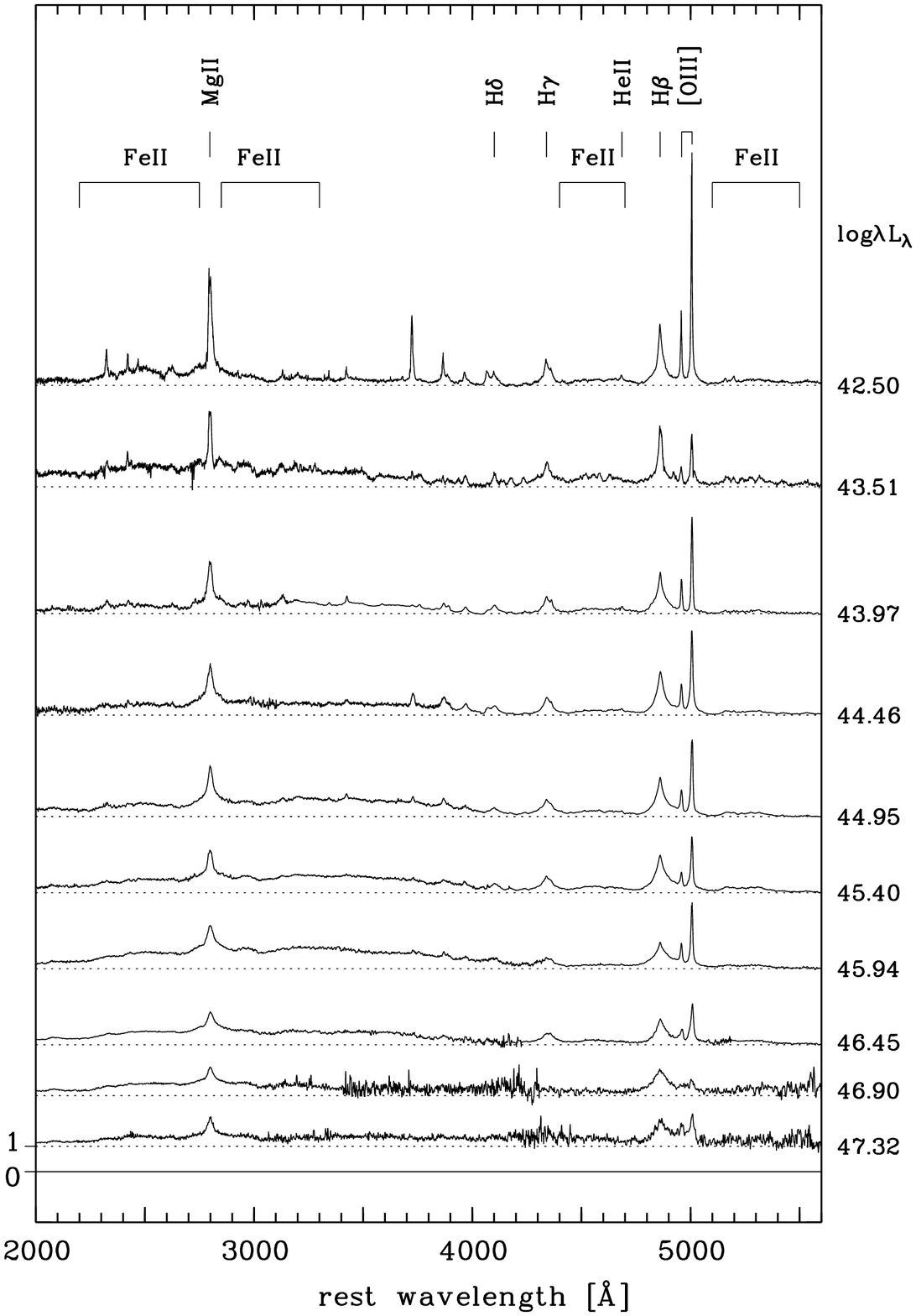}
\caption{Same as Fig.\,5, but the normalized composite spectra are shown at
        larger wavelengths.
\label{fig6}}
\end{figure}

\begin{figure}
%\plotone{beffpapf07afin.ps}{beffpapf07bfin.ps}
\plottwo{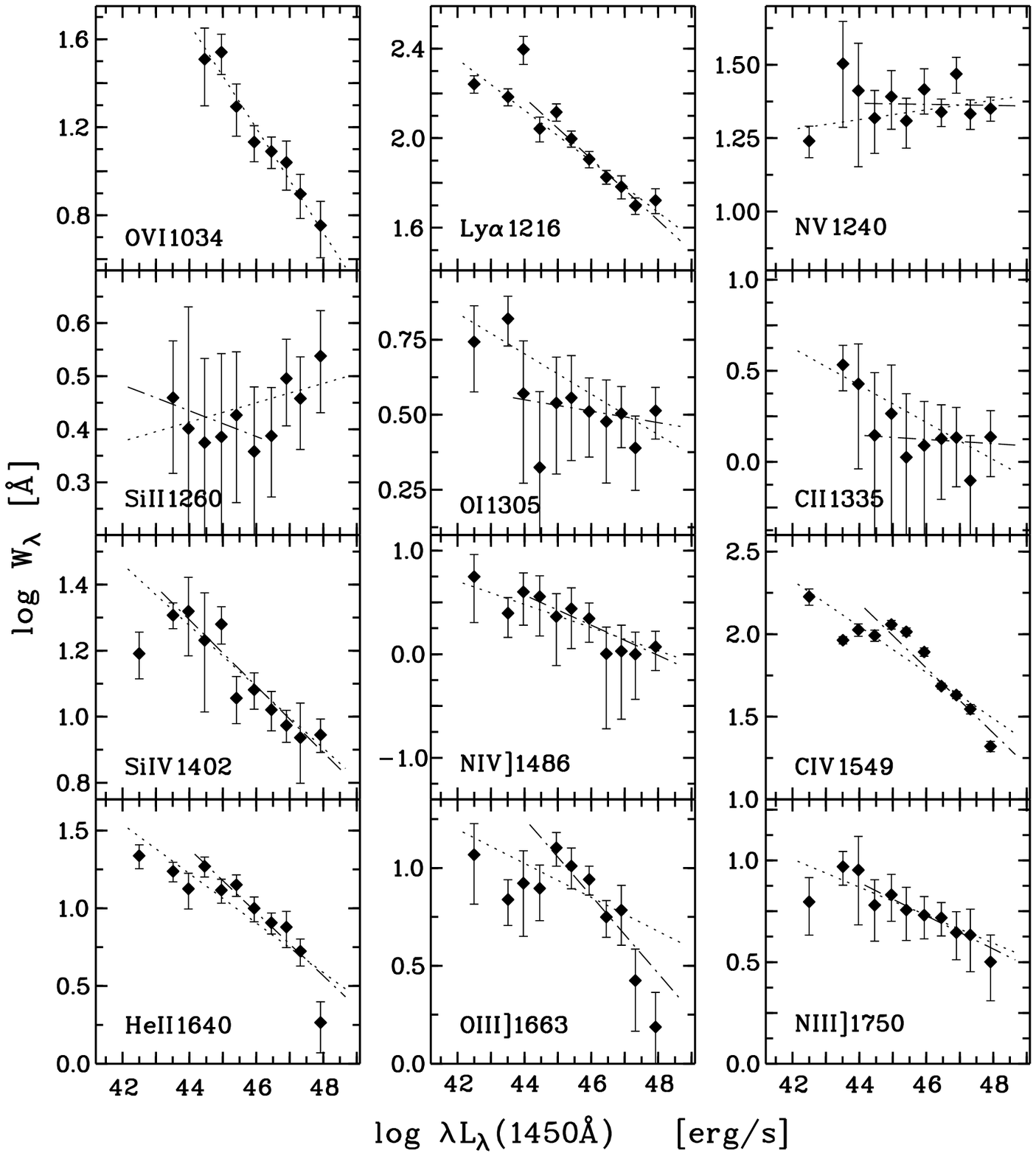}{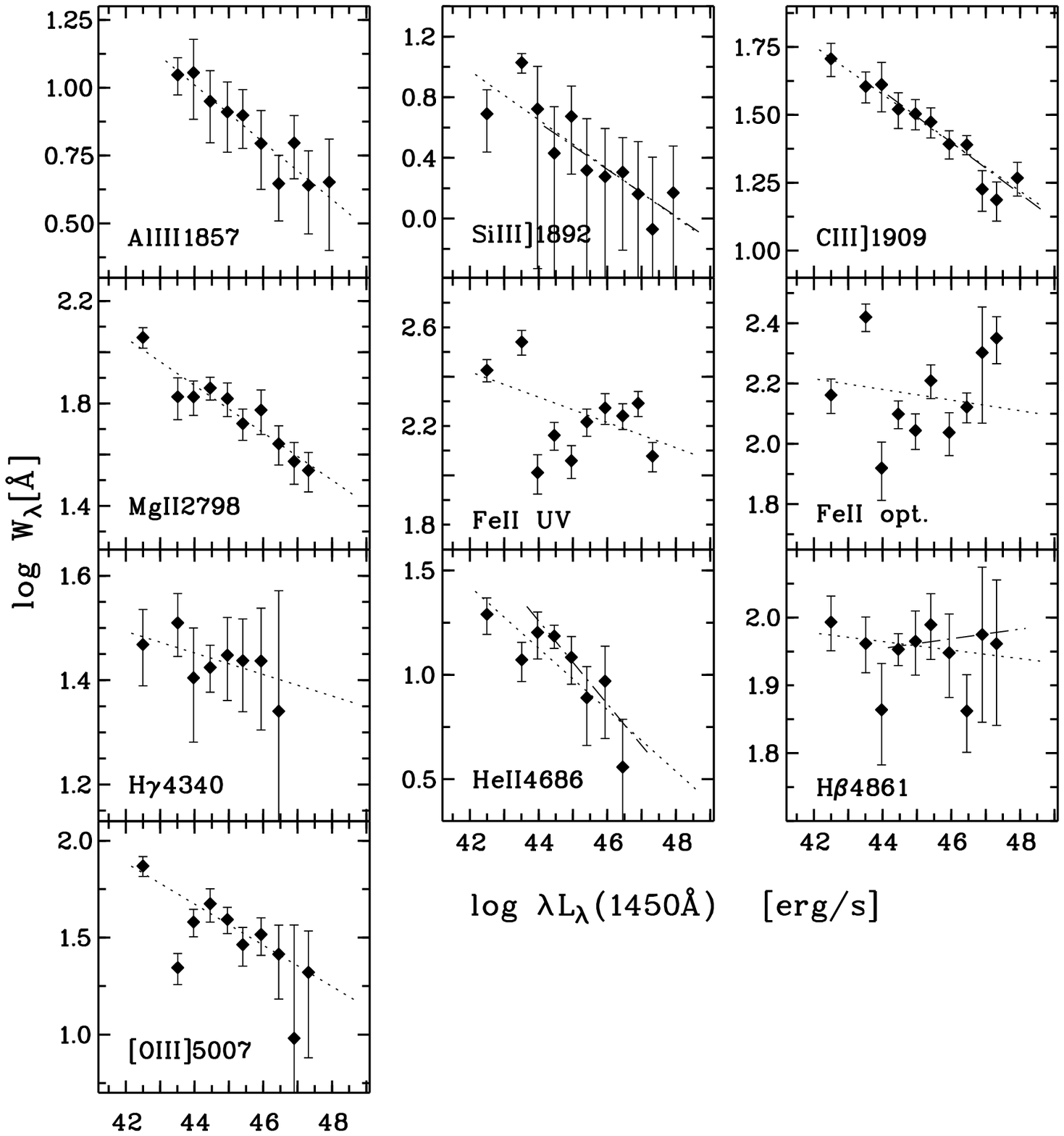}
\caption{Line equivalent widths, $W_\lambda$, as a function of increasing
         continuum luminosity $\lambda\,L_\lambda (1450{\rm \AA })$.
         We calculated linear fits to $W_\lambda $(L) for the entire
         luminosity range (dashed line), as well as a luminosties
         log\,$\lambda\,L_\lambda (1450{\rm \AA }) \geq 44$ (dashed-dotted 
         line).
\label{fig7}}
\end{figure}

%\begin{figure}
%%\plotone{beffpapf07bfin.ps}
%\plotone{f7b.eps}
%\caption{Same as Fig.\,7a, but for additional emission lines. 
%\label{fig7b}}
%\end{figure}

\begin{figure}
\plotone{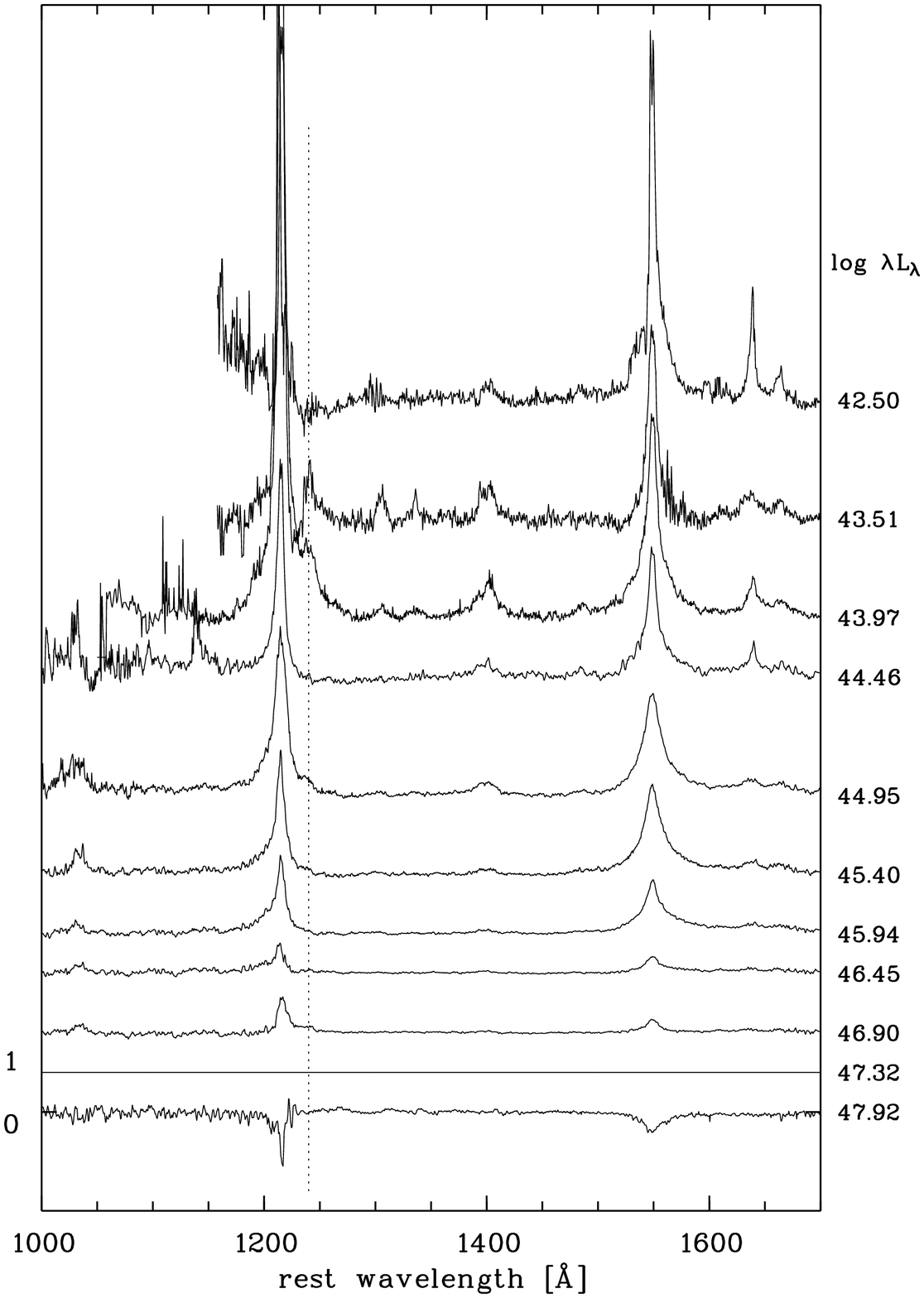}
\caption{Fig.\,8 --- Difference spectra with respect to the composite spectrum
            log\,$\lambda\,L_\lambda (1450{\rm \AA}) = 47.32$ are displayed
            to illustrate that $W_\lambda$(\nv $\lambda 1240$) is nearly
            constant, regardless how $W_\lambda $(\nv) is measured,
            while the other emission lines display a prominant BEff.
            The dotted line indicates the location of \nv $\lambda 1240$.
\label{fig8}}
\end{figure}

\begin{figure}
%\plotone{beffpapf08fin.ps}
\plotone{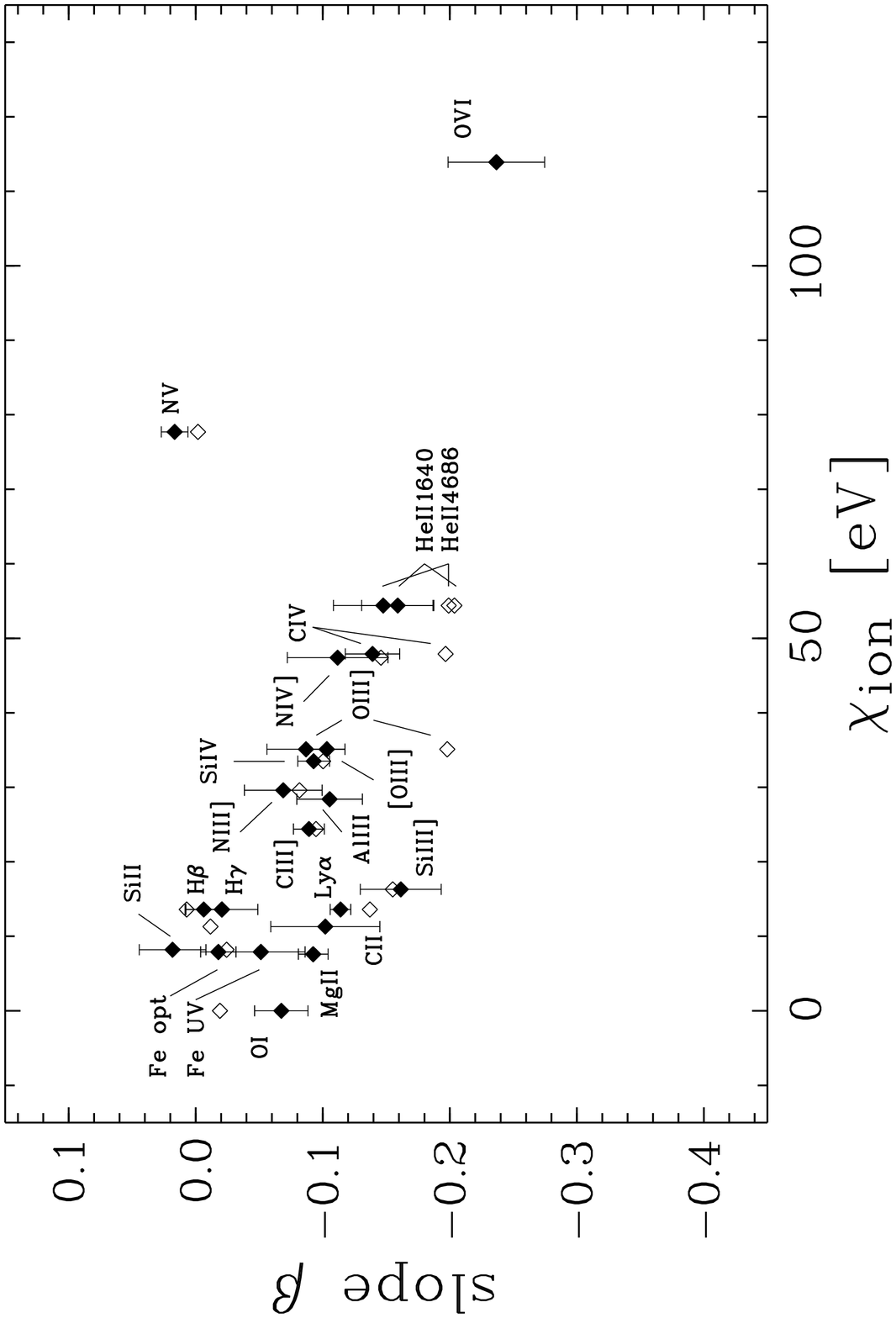}
\caption{The slope of the Baldwin Effect as displayed in Figs.\,7a and 7b as a 
         function of the ionization energy $\chi _{ion}$ needed to create rhe
         specfic ions. Different lines of the same ion 
         (\heii $\lambda 1640$, \heii $\lambda 4686$ and \oiii ]$\lambda 1663$,
         \ob ) show nearly identical slopes.
         The filled symbols represent the slopes based on the entire 
         luminosity range, corresponding to the dashed lines in Fig.\,7a. 
         The slopes of the Baldwin Effect for higher luminosities only 
         (log\,$\lambda\,L_\lambda (1450{\rm \AA }) \geq 44$) are plotted as 
         open symbols.
\label{fig9}}
\end{figure}

\clearpage

\clearpage

\begin{table}
\begin{center}
\caption{Slope $\beta $ of the Baldwin Effect for the individual emission 
         lines. A linear least square fit was calculated for the entire 
         luminosity range and for log\,$\lambda\,L_\lambda (1450{\rm \AA }) 
         \ga 44$; 
    log\,$\,W_\lambda = a + \beta ~ \log\,\lambda\,L_\lambda (1450{\rm \AA })$.
         In addition, for comparison theoretical predictions are given 
         (Korista et al.\,1998).
         \label{tbl-1}}
\begin{tabular}{lrccc}
\tableline\tableline
line &$\chi _{ion}$&entire $\lambda\,L_\lambda (1450{\rm \AA })$-range&
                    log\,$\lambda\,L_\lambda (1450{\rm \AA })\ga 44$&
                    Korista et al.(1998)\\
     & [eV]&$\beta $&$\beta$&$\beta $\\
%\multicolumn{1}{c}{$P$\tablenotemark{a}} & $P R_{maj}$ & $P R_{min}$ &
\tableline
\ovi $\lambda 1034$   &$113.9$&$-0.24 \pm 0.04$&$-0.24 \pm 0.04$&$-0.22$\\
\nv  $\lambda 1240$   &$ 77.7$&$+0.01 \pm 0.01$&$-0.00 \pm 0.02$&$-0.05$\\
\heii $\lambda 1640$  &$ 54.4$&$-0.16 \pm 0.03$&$-0.20 \pm 0.04$&$-0.17$\\
\heii $\lambda 4686$  &$ 54.4$&$-0.15 \pm 0.04$&$-0.20 \pm 0.09$& \nodata \\ 
\civ $\lambda 1549$   &$ 47.9$&$-0.14 \pm 0.02$&$-0.20 \pm 0.03$&$-0.20$\\
\niv ]$\lambda 1486$  &$ 47.4$&$-0.11 \pm 0.04$&$-0.14 \pm 0.08$&$-0.01$\\
\oiii ]$\lambda 1663$ &$ 35.1$&$-0.09 \pm 0.03$&$-0.20 \pm 0.05$&$-0.12$\\
\ob                   &$ 35.1$&$-0.10 \pm 0.02$&   \nodata      & \nodata \\
\siiv $\lambda 1402$  &$ 33.5$&$-0.09 \pm 0.01$&$-0.10 \pm 0.02$&$-0.13$\\
\niii ]$\lambda 1750$ &$ 29.6$&$-0.07 \pm 0.03$&$-0.08 \pm 0.04$&$+0.09$\\
\aliii $\lambda 1857$ &$ 28.4$&$-0.11 \pm 0.03$&$-0.11 \pm 0.03$&$-0.04$\\
\ciii ]$\lambda 1909$ &$ 24.4$&$-0.09 \pm 0.01$&$-0.09 \pm 0.02$&$-0.08$\\
\siiii ]$\lambda 1892$&$ 16.3$&$-0.16 \pm 0.03$&$-0.16 \pm 0.12$&$-0.06$\\
\lya $\lambda 1216$   &$ 13.6$&$-0.11 \pm 0.01$&$-0.14 \pm 0.02$&$-0.10$\\
\hgamma $\lambda 4340$&$ 13.6$&$-0.02 \pm 0.03$&$-0.02 \pm 0.03$& \nodata \\ 
\hbeta $\lambda 4861$ &$ 13.6$&$-0.01 \pm 0.01$&$+0.01 \pm 0.03$& \nodata \\
\cii $\lambda 1335$   &$ 11.3$&$-0.10 \pm 0.04$&$-0.01 \pm 0.09$&$+0.00$\\
\sizwo $\lambda 1260$ &$  8.2$&$+0.02 \pm 0.03$&$-0.01 \pm 0.09$& \nodata \\
\feii UV              &$  7.9$&$-0.05 \pm 0.03$&  \nodata       & \nodata \\
\feii opt             &$  7.9$&$-0.02 \pm 0.03$&  \nodata       & \nodata \\ 
\mgii $\lambda 2798$  &$  7.6$&$-0.09 \pm 0.01$&$-0.09 \pm 0.01$&$-0.09$\\ 
\oi $\lambda 1305$    &$  0.0$&$-0.07 \pm 0.02$&$-0.02 \pm 0.04$&$-0.05$\\
\tableline
\end{tabular}
\end{center}
\end{table}

\clearpage

%% The following command ends your manuscript. LaTeX will ignore any text

\end{document}